\documentclass[fleqn,usenatbib]{mnras}

\usepackage[T1]{fontenc}

\usepackage{graphicx}	
\usepackage{amsmath}	
\usepackage{amssymb}	
\usepackage{xcolor}
\usepackage{hyperref}
\usepackage{wasysym}
 \usepackage{float}
\usepackage{ulem}
\usepackage{comment}
\usepackage{mdframed}
\usepackage{newtxtext,newtxmath}
\usepackage{soul}
\usepackage{cleveref}

\newcommand{\s}[1]{\textrm{#1}}

\def\apj {ApJ}

\def\apjs {ApJS}
\def\aj {AJ}
\def\aap {A\&A}
\def\mnras {MNRAS}

\title[Deep Learning DM determination]{Determining the Dark Matter distribution in simulated galaxies with Deep Learning
}
\author[M. de los Rios et al.]{
Mart\'in de los Rios,$^{1,2,3}$
Mihael Peta\v{c},$^{4,5}$
Bryan Zaldivar,$^{6}$
Nina R. Bonaventura,$^{7}$
Francesca Calore$^{8}$, and
\newauthor
Fabio Iocco $^{9}$\thanks{fabio.iocco.unina@gmail.com}
\\
\\
$^{1}$ICTP South American Institute for Fundamental Research \& Instituto de F\'isica Te\'orica, Universidade Estadual Paulista, 01140-070, S\~ao Paulo-SP, Brazil\\
$^{2}$Departamento de F\'isica Te\'orica, Universidad Aut\'onoma de Madrid, 28049 Madrid, Spain\\
$^{3}$Instituto de F\'isica Te\'orica, UAM-CSIC, c/ Nicolás Cabrera 13-15, Universidad Autónoma de Madrid, Cantoblanco, Madrid 28049, Spain\\
$^{4}$Center for Astrophysics and Cosmology (CAC) of University of Nova Gorica, Vipavska 11c, 5270 Ajdovščina, Slovenia\\
$^{5}$Laboratoire Univers et Particules de Montpellier (LUPM), Universit\'e de Montpellier (UMR-5299) \& CNRS, Place Eug\`ene Bataillon, F-34095\\ Montpellier Cedex 05, France\\
$^{6}$Institute of Corpuscular Physics (IFIC), University of Valencia and CSIC, Spain\\
$^{7}$Cosmic Dawn Center, Niels Bohr Institute, University of Copenhagen, Denmark\\
$^{8}$Univ. Grenoble Alpes, Univ. Savoie Mont Blanc, CNRS, LAPTh, F-74940 Annecy, France\\
$^{9}$Dipartimento di Fisica "Ettore Pancini" Università degli studi di Napoli ``Federico II'' \& INFN sezione di Napoli, Italy
}

\date{Accepted XXX. Received YYY; in original form ZZZ}

\pubyear{2023}

\begin{document}
\label{firstpage}
\pagerange{\pageref{firstpage}--\pageref{lastpage}}
\maketitle

\begin{abstract}
We present a novel method of inferring the Dark Matter (DM) content and spatial distribution within galaxies, using convolutional neural networks (CNNs) trained within state-of-the-art  hydrodynamical simulations (Illustris--TNG100).
 Within the controlled environment of the simulation, 
the framework we have developed is capable of inferring the DM mass distribution within galaxies of mass $\sim 10^{11}$ -- $10^{13} \, M_\odot$ 
from the gravitationally baryon-dominated internal regions to the DM-rich, baryon-depleted outskirts of the galaxies, with a mean absolute error always below $\approx0.25$ when using photometrical and spectroscopic information.
With respect to traditional methods, the one presented here also possesses the advantages of not relying on a pre-assigned shape for the DM distribution, to be applicable to galaxies not necessarily in isolation, and to perform very well even in the absence of spectroscopic observations. 

\end{abstract}




\section{Introduction}
\label{sec:introduction}

\par The prediction of the formation of galaxies from the initial perturbations of matter seen in the Cosmic Microwave Background (CMB) within a self-consistent framework, is one remarkable success of the $\Lambda$ Cold Dark Matter ($\Lambda$CDM) paradigm.
In particular, the presence of a gravitationally active, otherwise inert, component of matter --whose intimate nature is currently unknown-- dubbed ``Dark Matter'' (DM), is what is believed to allow the density perturbations observed in the CMB to grow into large scale structures, providing the leading ``gravitational texture'' to the fabric of the Universe we observe today.

\par Achieving a full theoretical understanding of the evolution and growth of density perturbations is a very challenging task; semi--analytical solutions have been devised \citep{Press:1973iz, Sheth:1999su}, which are able to capture the growth of the power spectrum on large and medium scales, down to the size of bound objects.
On smaller scales, the distribution of matter within galaxies themselves is typically learned through the solution of the equations describing the growth of an inert component, coupled to that of an active component, which requires numerical treatment.
Without any other assumption on the nature of DM other than the lack of any particle interaction with  ordinary matter, one can solve the equations for the growth of the primordial density perturbations containing both  ordinary matter (hereafter ``baryons'') and the DM, into the potential wells that will later become galaxies and galaxy clusters.
This class of methods, generically known as ``hydrodynamic simulations'', has been developed in the past two decades by several groups which have extensively tested different techniques to solve the evolution of the gaseous component of the Universe using computational methods from fluid dynamics, which have proven successful in reproducing a number of properties of visible galaxies.
We list here only a few of these simulations, not attempting an exhaustive review of the field: the Illustris and the Illustris-TNG projects~\citep{Vogelsberger:2014kha, Vogelsberger:2014dza,Pillepich:2017jle,Nelson:2017cxy}, the EAGLE and APOSTOLE projects~\citep{2015MNRAS.450.1937C,Schaye:2014tpa,Sawala:2015cdf,2016MNRAS.457..844F}, the NIHAO simulations suite~\citep{Wang:2015jpa}, the FIRE project~\citep{Hopkins:2013vha}, and the ERIS simulation~\citep{2011ApJ...742...76G}.
These simulations ultimately aim to satisfy the precise observational constraints that are (and will be) offered by large-scale-structure surveys including SDSS~\citep{sdssIV}, 2DF~\citep{2df}, DEEP2~\citep{deep2}, and CANDELS~\citep{candels}, and upcoming projects such as Vera Rubin (also known as LSST)~\citep{lsst}.

\par Observationally, the DM distribution in objects of diverse masses and spatial extent is inferred through a host of different methods, ranging from the use of gravitational lensing in galaxy clusters,  
to the solution of inverted Jeans equations for dwarf galaxies, 
through to the most famous `rotation curve' technique for disk galaxies. 

All of the above methods are based on assumptions known to hold in what can be considered a `controlled environment',
for instance and only as a non-comprehensive example: the fact that the stellar disk is rotationally supported in disk galaxies for rotation curve methods, or some simplifying assumptions on the shape of, e.g., the anisotropy velocity parameter for reconstructing the DM potential in dwarf galaxies~\citep{Strigari:2018utn}. These assumptions are known to eventually break down and introduce unavoidable systematics for which one must account (see, e.g., \cite{Ullio:2016kvy}). 

To overcome limitations in the DM reconstruction related to the lack of a controlled environment where strict validity/accuracy tests may be performed, a method to recover the actual DM content of a galaxy (or a set of galaxies), trusted and validated through an accuracy test, possibly in a controlled environment, is required. 
The existence of numerical simulations, described above, offers an ideal test-bed to address this problem. In such a controlled environment, one is in control of both the visible and invisible components of the Universe (within the corresponding simulation's cosmological box), and is also able to produce images of the visible Universe that emulate the ones actually observed with existing instruments.
This offers the unique chance to perform controlled tests of the methods devised, and/or to develop brand new ones, based on entirely novel technologies.
\par Machine learning (hereafter, ML) techniques have proven to be powerful tools for classification and regression tasks in astronomy and astrophysics (e.g.,~\cite{Hezaveh:2017sht, 2019MNRAS.488..991P, Chianese:2019ifk,Necib:2019zka}), 
as they look for correlations between the input variables and the output variables one wants to infer.
It is important to remark that, in order to achieve this goal, it is necessary to have a reliable data set for which the input variables are known (in the present analysis, photometric images and spectroscopy of galaxies), as well as the output variables  (in the present analysis, the underlying DM distribution).

 In the present work, we rely on machine learning algorithms, most notably {\it deep learning}, to develop a new method for recovering the DM distribution within galaxies from photometric and spectroscopic observations.
In order to achieve this goal, we train the machine learning algorithms in an environment where the underlying `{\it truth}' is known, and on simulated images resembling those obtained from real observatories with a high degree of approximation.
{\it Machines} can thus properly be trained in the datacubes of numerical simulations, and applied to images of the real, external, environment. 

Here we list the key elements of our approach:
\begin{itemize}
\item We use the Illustris-TNG100 simulation suite, and select galaxies  with total stellar mass in the range of $M_\star \in [10^{10} \; M_\odot, 10^{12} \; M_\odot]$.
\item We augment the mock galaxy sample with simulations of realistic telescope observables (such as HI data cubes and velocity maps) to train our {\it machines} in a controlled environment that resembles the real Universe. 
\item We thoroughly test deep learning network architectures to identify the best one suited for our purposes. 
\item With our {\it machines}, we reconstruct the DM density distribution without relying on assumptions of a specific shape for the profile, contrary to other methods looking only at the total mass, or mass enclosed within a particular radius.
\item Our approach is not limited to isolated rotation-supported systems like the rotation curve method, but is applicable to the whole range of galaxies in the local Universe, including those undergoing mergers, turbulent star formation, etc.
\item While the rotation curve method requires accurate photometric {\it and} spectroscopic observations to determine the DM content of a galaxy, our approach remarkably achieves very good performance with photometry only.
\end{itemize}

We describe in detail all of these elements in this paper, organized as follows:
In Section \ref{sec:simulations}, we describe the set of cosmological numerical simulations that we adopt in order to create our sample of galaxies; in Section~\ref{sec:mocks} we describe the creation of mock images that resemble (for the galaxies extracted in the simulation) the observational properties of real observations of galaxies; in Section~\ref{sec:networks} we describe the machine learning algorithms (to which we will refer to as either {\tt architectures} or {\tt machines}) that we train inside the simulation space; in Section~\ref{sec:results} we present and discuss our results. We finally draw our conclusions in Section~\ref{sec:conclusion}.

\section{The simulations}
\label{sec:simulations}

\par The first numerical solutions of structure formation in a cosmological environment (hereafter, referred to as `cosmological simulations') described the formation and growth of DM--only structures, while subsequent developments improved their realism by including various aspects of baryonic physics. Due to the vastly different scales of various processes that govern the behaviour of baryons, this turns out to be a very demanding task. Within the past decades, sophisticated simulation frameworks have been developed, which rely on the modelling of collisional baryonic matter as a fluid, and are therefore referred to as \textit{hydrodynamical simulations}. The baryonic physics below the resolution limit is typically handled through various prescriptions that are calibrated on a broad range of observations  -- for a review see, e.g.,~\cite{Pillepich:2017jle}. This approach, aided by the rapid improvement of computational capabilities, has led to the development of highly reliable simulations, which can generate spectacularly accurate analogues of our Universe and manage to reproduce a wide range of empirical relations that are known to exist in real galaxies.

\par In this work we explore a novel simulation-based technique for determining the DM content of the observed galaxies. Within the simulations, one has a complete knowledge of all the components of the Universe (gas, dust, stars, accreting black holes, and Dark Matter), thus being able to create  mock observations of galaxies and pair them to the corresponding DM mass profiles. This setup enables us to use supervised machine learning techniques to capture the mapping between the observational data and the underlying DM distribution. Naturally, the success of the outlined approach crucially relies on the assumption that the simulations are representative of the real Universe. As we discuss in the following section, the IllustrisTNG simulations, on which we base our work, indeed provide an excellent agreement with a broad range of observations, which go far beyond the quantities that were used in the calibration of sub-grid physics.

\subsection{The Illustris TNG simulations}

The IllustrisTNG simulations (\cite{Marinacci2018, Naiman2018, Nelson2018, Pillepich2018, Springel2018}) are built upon the success of their predecessors, namely the original Illustris simulations. While the latter did not include magnetic fields, and used less refined implementations of AGN feedback, stellar mass yields and galaxy-wide winds, they were one of the first simulations to successfully reproduce observed galaxy morphologies~\citep{vogelsberger_2014}, besides several other global properties; these include such properties as the halo mass function, galaxy stellar mass function, luminosity function, Tully-Fisher relation, as well as a fair approximation to the star formation rate (SFR) and its evolution~\citep{vogelsberger_2014_2}. However, the Illustris runs also suffered from several shortcomings, e.g. in a mismatch in stellar ages of small-mass galaxies and in the realm of quenching of massive galaxies (for a complete review of scientific remarks and cautions see~\citep{nelson_2015}), which motivated the development of The Next Generation (TNG) simulation suit. The latter notably improved on the aforementioned shortcomings~\citep{2018MNRAS.473.4077P} thanks to the refined implementation of feedback, as well as the implementation of magnetic fields. Additionally, the IllustrisTNG simulations have also proven to perform better when subjected to more stringent tests, such as the matching against the observed galactic size-mass ($R_e - M_\star$) relation~\citep{2018MNRAS.474.3976G} and the evolution of the SFR~\citep{2019MNRAS.485.4817D}, although some discrepancies regarding the latter still exist at $z \gtrsim 1$~\cite{donnari_2019}. Even more importantly for our work, the IllustrisTNG suite was shown to result in remarkable agreement between simulated and observed galaxy morphologies~\citep{2019MNRAS.489.1859H}; allow for the generation of representative mock observations for Pan-STARRS~\citep{chambers2019panstarrs1} and the SDSS~\cite{sdssIV} survey~\citep{nelson_2015,RodriguezGomez2019} ; and lead to rotation curves that are similar to the ones observed in Milky Way-like galaxies~\citep{lovell_2018,marasco_2020}.

In order to build our data-set we use the TNG100-1 simulation run. Like other TNG simulations, it self-consistently follows the formation and evolution of galaxies and their environments for $100$ snapshots from $z=127$ to $z=0$ in a $\Lambda$CDM cosmology ($\Omega_{m} = 0.3089$, $\Omega_{b} = 0.0486$, $\Omega_{\Lambda} = 0.6911$, $\sigma_{8} = 0.8159$, $n_{s} = 0.9667$ and $h = 0.6774$ \citep{Planck2016}). This simulation is embedded in a periodic box with sides of $75h^{-1} \approx 110.7 Mpc$ and $2 \times 1820^{3}$ resolution elements, which translates to an average mass of the baryonic resolution elements of $1.39 \times 10^{6}M_{\odot}$ and DM particles with mass $7.5 \times 10^6 \, M_\odot$ \citep{RodriguezGomez2019}.

\subsection{Selection of galaxies}
\label{sec:simulations_selection}

For the purposes of this work, we use a subset of galaxies from the TNG100-1 simulation that were selected according to criteria motivated by the above described consistency checks of the TNG suite, as well as the availability of the real-world observations. In particular, we restrict our attention to {\tt snapshot 99}, corresponding to redshift $z = 0$, since in this pilot study we  focus on local galaxies for which reliable independent mass estimates are available (e.g., the SPARC catalogue~\citep{Lelli2016}). We further restrict our attention to galaxies that have  stellar masses in the range of $M_\star \in 10^{10} \, M_\odot$ -  $10^{12} \, M_\odot$, as these are well-resolved by the simulation and pass a broad range of tests discussed in the previous section. Furthermore, this stellar-mass range is also well covered by the existing photometric and interferometric surveys. Apart from the stellar mass, we also impose a cut on SFR in order to select only objects with a SFR exceeding $0.1 \; M_\odot / \s{yr}$  to avoid quenched galaxies, which have been reported to possess notable discrepancies with respect to the observations~\citep{2019MNRAS.489.1859H}. Furthermore, the cut on SFR also assures that the objects have a sufficient amount of gas for creating the corresponding mock interferometric images. Finally, in addition to the cuts on $M_\star$ and SFR, we also ensure that the selected galaxies are central galaxies (i.e., their SubhaloParent = 0, which implies they have no parent in the hierarchical structure resolved by the Subfind~\cite{2001MNRAS.328..726S} algorithm) and that they have been identified as true galaxies by the Illustris collaboration (i.e., have SubhaloFlag set to 1, which assures us that they are proper galaxies of cosmological origin). It is to be noted that the selection criteria described above do not narrow down our sample of galaxies to a specific morphological type, but includes all types of objects that can be found in the TNG100-1 simulation and pass our stellar mass and SFR cuts. Similarly, the selected galaxies are not necessarily isolated, since we only demand that they are the central object of the corresponding friend-of-friends group identified by the Subfind algorithm. The summary of our selection criteria can be found in Table~\ref{tab:selection_criteria}.

\begin{table}
    \centering
    \begin{tabular}{c|c}
        Property & Criterium \\ \hline
        Simulation snapshot & $99$ $(z = 0)$ \\
        Stellar mass & $10^{10} \; M_\odot \leq M_\star \leq 10^{12} \; M_\odot$ \\
        Star formation rate & $\s{SFR} \geq 0.1$ $M_\odot / \s{yr}$ \\
        Central galaxy & SubhaloParent = 0 \\
        Cosmological origin & SubhaloFlag = 1
    \end{tabular}
    \caption{Summary of the criteria used in selecting galaxies from the simulation.}
    \label{tab:selection_criteria}
\end{table}

\section{Creation of the mock observations}
\label{sec:mocks}

\par The existence of reliable and representative training data is a crucial precondition for the applicability of supervised machine learning tools. As explained in the previous section, modern hydrodynamical simulations provide us with a realistic representation of the Universe and, in particular, its galaxy populations.
In combination with state-of-the-art image generation tools, this give us a unique opportunity to create large and realistic sets of mock observations of galaxies. Furthermore, since the exact underlying DM distribution can be determined from the simulations, they provide us with all the required ingredients to apply the standard supervised machine learning techniques.

The creation of realistic mock observations from hydrodynamical simulations has been insofar explored by a number of works~\citep{Nelson2018,RodriguezGomez2019}. Therefore, we chose to follow the prescriptions therein, as they have been thoroughly tested and shown to non-trivially reproduce a number of observational properties. More concretely, throughout the rest of this work we will rely on mock photometric images of the Sloan Digital Sky Survey (SDSS)~\citep{sdssIV} and HI interferometry mimicking the characteristics of Karl G. Jansky Very Large Array (VLA) radio observatory~\citep{VLA_2020}. Such a combination of photometric images and gas kinematics has been frequently used to constrain the DM content in local spiral galaxies, see, e.g.,~\cite{Lelli2016}.

As detailed in Section~\ref{sec:simulations_selection}, we apply several selection cuts to the full sample of galaxies provided by the TNG100-1 simulation. In particular, our subset corresponds to local central galaxies with stellar masses in the range of $M_\star \in 10^{10} \, M_\odot$ - $10^{12} \, M_\odot$ and SFR $\geq 0.1 \; M_\odot / \s{yr}$. After applying these selection criteria, we are left with approximately 2000 individual objects. To increase the size of the data set, we generate 3 distinct realisations of photometric images and interferometric data for each galaxy in our sub-sample, with each realisation having a different randomly selected orientation (i.e. line-of-sight axis) and distance to the object, which is selected uniformly from the interval $D \in 10 \, \s{Mpc}$ - $20 \, \s{Mpc}$. All the mock observations are fixed to cover $17.2' \times 17.2'$ of the sky, have a resolution of $128 \times 128$ pixels and are centred on the most gravitationally bound particle belonging to the object.

In the following section, we first summarise the procedure for generating mock photometric images in the five SDSS wavebands. Subsequently, we describe the method used for creating HI intensity and velocity maps, which were obtained from mock interferometric data cubes resembling the observations of VLA.

\subsection{Creation of mock photometric images}
\label{sec:mocks_photometry}

For the creation of mock SDSS images we follow the procedure established by~\cite{RodriguezGomez2019}. It relies on the radiative transfer code \texttt{SKIRT} \citep{skirt,skirt9}, which emulates the stellar emissions and subsequent light-ray propagation to the observer, taking into account the absorption and re-emission by dust, for a given cut-out of particles from the hydrodynamical simulation. It allows us to directly produces the idealised mock images for the five SDSS broadband filters, while we additionally add the seeing and sky brightness corrections to each of the simulated bands according to the characteristics reported by the SDSS collaboration.

The first step in producing mock photometric images for a simulated galaxy, is obtaining the corresponding cut-outs for stellar and gas particles from the simulation. For a given object, defined by the friends-of-friends (FoF) algorithm used in the pre-processing carried out by the IllustrisTNG collaboration~\citep{Springel2018}, we extract all the stellar and gas particles which lie within a sphere centred on the corresponding minimum of the gravitational potential.~\footnote{This procedure selects also particles that are not necessarily assigned to galaxy in question by the FoF algorithm, but might nonetheless contribute to the propagation of light along the line-of-sight in the vicinity of the object (e.g. satellite galaxies or other nearby neighbours).} The radius of the sphere is chosen according to our prescribed field of view, i.e. covering $17.2'$ at a previously selected random distance $D$. After extracting the corresponding set of particles, we apply a random 3D rotation to obtain the final cut-out which is to be used as an input for the \texttt{SKIRT} code.

Within the \texttt{SKIRT} framework, each imported star particle is treated as single coeval stellar population. Following the approach of~\cite{RodriguezGomez2019}, the spectral energy distributions (SEDs) of stellar particles older than $10 \, \s{Myr}$ are modelled with the Bruzual \& Charlot~\citep{Bruzual2003} population synthesis code, for which the initial mass, metallicity and age are provided as additional inputs, obtained directly from the simulation. The stellar particles younger than $10 \, \s{Myr}$ are treated as star-bursting regions and their SEDs are modelled with the MAPPINGS-III photo-ionization code \citep{Groves2008}. For the latter, a constant SFR over the last $10 \,  \s{Myr}$ is assumed, along with a compactness parameter $\log_{10} \mathcal{C} = 5$, interstellar medium pressure of $\log_{10}[(P_{0}/k_{B})/\s{cm}^{-3}K] = 5$ and a cloud covering factor of $f_\s{PDR} = 0.2$. The corresponding metallicity values were again obtained directly from the simulation. The SEDs of both, old stellar populations and star-forming regions, are sampled using 1000 logarithmic wavelength bins in the range between 0.09 to 100 $\mu \s{m}$. 

For performing the radiative transfer computation, \texttt{SKIRT} uses a self-consistent description of the gas distribution with the one used throughout in the simulation. In particular, it reconstitutes the original Voronoi tessellation that underpins the adaptive moving mesh \texttt{AREPO} code, upon which the IllustrisTNG suite is built. Following the convention of~\cite{RodriguezGomez2019}, we assume that dust follows the distribution of star-forming gas and has a constant dust-to-metal mass ratio of $0.3$. The dust composition is modelled using the multicomponent dust mix of \cite{Zubko2004}, which consists of graphite, silicate and polycyclic aromatic hydrocarbon grains. For the dust emissivity we assume a modified blackbody spectrum, which relies on local thermodynamic equilibrium, and track it over 1000 logarithmic bins spanned between 0.09 and 100 $\mu \s{m}$.

The above settings are used to run a Monte-Carlo simulation of radiative transfer for $10^7$ photon packets, which is significantly higher than what was used in~\cite{RodriguezGomez2019} -- the reason for this difference is the fact that we assume the sources are significantly closer and hence better resolved. Similarly, we use a frame instrument with 1000 logarithmically spaced bins in the range between 0.01 and 3.7 $\mu \s{m}$, which leads to significantly better spectral resolution. Finally, the resulting data cube is convolved with the five SDSS \textit{ugriz} broadband filters to obtain the corresponding images of the optical stellar light emission.

\begin{figure*}
	\includegraphics[width=\linewidth]{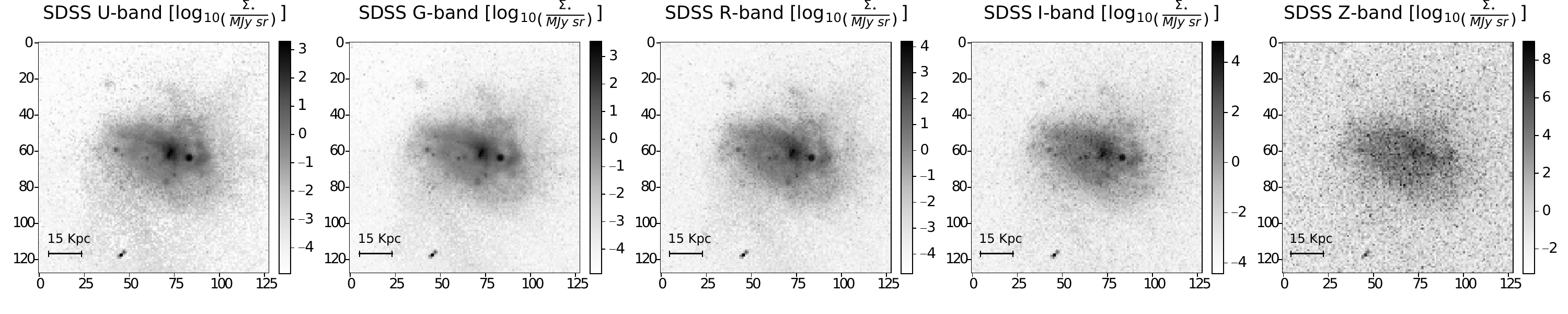}
	\includegraphics[width=\linewidth]{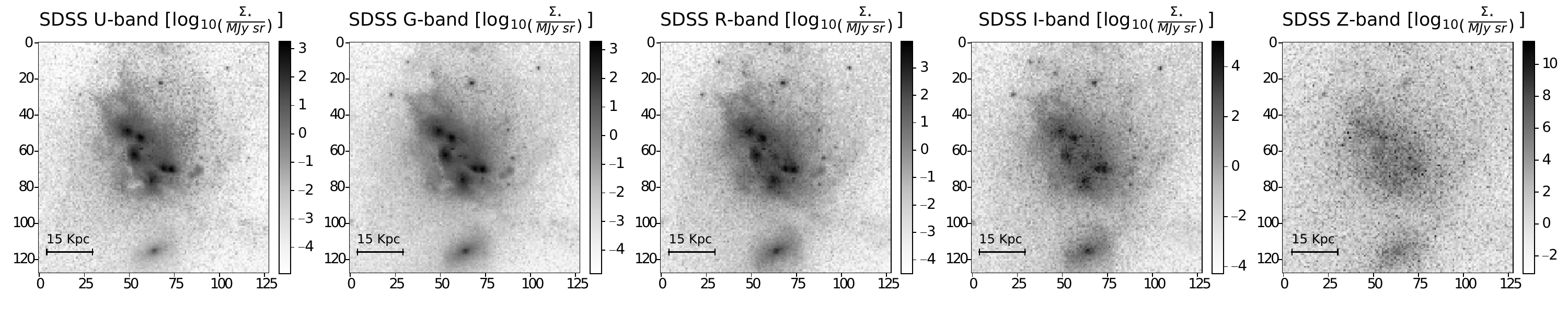}
	\caption{Examples of the photometric mock SDSS  images obtained following the procedure described in Sec. \ref{sec:mocks_photometry} for two random galaxies (subhaloID = 60744 in the top panel and subhaloID = 108013 in the bottom panel).}
	\label{fig:SDSS}
\end{figure*}

It is worth emphasizing that this procedure yields idealized photometric images for the simulated galaxies. To produce realistic mock observations, one needs to, at a minimum, add the `blurring' effects to the simulated light emission that would result if it passed through the optics of a telescope, as well as the contaminating contribution to the observed emission from the sky background. Therefore, we convolve each image with a Gaussian kernel filter appropriately sized to match the width of the point spread function (PSF) in each of the SDSS wavebands, which we adopt from the SDSS DR16~\footnote{\href{https://www.sdss.org/dr16/imaging/other_info/}{\url{https://www.sdss.org/dr16/imaging/other_info/}}}. 
Similarly, we add the sky background contribution in the form of random Gaussian noise to each image pixel, where the appropriate mean and variance values for each band were again taken from SDSS DR16$^2$. The adopted PSF and median sky brightness values are contained in Table~\ref{tab:obs_corrections}. Examples of the final output images are shown in Figure~\ref{fig:SDSS}.

\begin{table}
	\centering
	\begin{tabular}{c|c|c}
		Filter & Seeing PSF [arcsec] & Sky brightness [mag /arcsec$^2$] \\ \hline
		u & 1.53 & 22.01 \\
		g & 1.44 & 21.84 \\
		r & 1.32 & 20.84 \\
		i & 1.26 & 20.16 \\
		z & 1.29 & 18.96
	\end{tabular}
	\caption{Seeing PSF and median sky brightness used in the creation of SDSS images. The values are adopted from SDSS DR16$^2$.}
	\label{tab:obs_corrections}
\end{table}

\subsection{Creation of HI intensity and velocity maps}
\label{sec:mocks_interferometry}

For the creation of mock HI observations, we follow the procedure described in~\cite{martini}, which is conveniently implemented in the \texttt{MARTINI} code~\footnote{\href{https://github.com/kyleaoman/martini}{https://github.com/kyleaoman/martini}}. This code allows for the creation of synthetic resolved HI line observations (i.e. data cubes) directly from the snapshot of a hydrodynamic simulation. It provides a broad range of features, ranging from spectral modelling to various observational effects, such as noise contamination and beam width corrections.

Since the support for the IllustrisTNG suite is pre-built within the \texttt{MARTINI} software, it allows us to easily obtain HI data cubes from the $z=0$ snapshot of the TNG100-1 simulation. For particle smoothing, we use the cubic spline kernel, which is a standard choice and, in our initial experiments, provided the best numerical stability. For the spectral dimension of the data cube, we use 64 channels with spectral resolution of $5 \, \s{km/s}$, which roughly matches the properties of the THINGS survey~\citep{things} obtained with the VLA radio telescope. We also use the corresponding beam properties, namely a Gaussian PSF with a full width at half maximum (FWHM) of $10''$, truncated at $3 \times$FWHM, and a Gaussian noise model with a root mean square value of $5 \times 10^{-6} \, \s{Jy/arcsec}^{2}$. Through the data source settings we removed the systemic velocities of the objects, so the resulting data cubes were produced in the appropriate galactic rest frames. For the distances to the objects, we used the same randomly generated values of $D$ as in the case of the photometric images. Furthermore, to obtain the identical orientations of the galaxies, we performed a rotation which aligned the particles extracted from the simulation box to the same line-of-sight direction as was used in the creation of mock SDSS images.

\begin{figure*}
	\includegraphics[width=\linewidth]{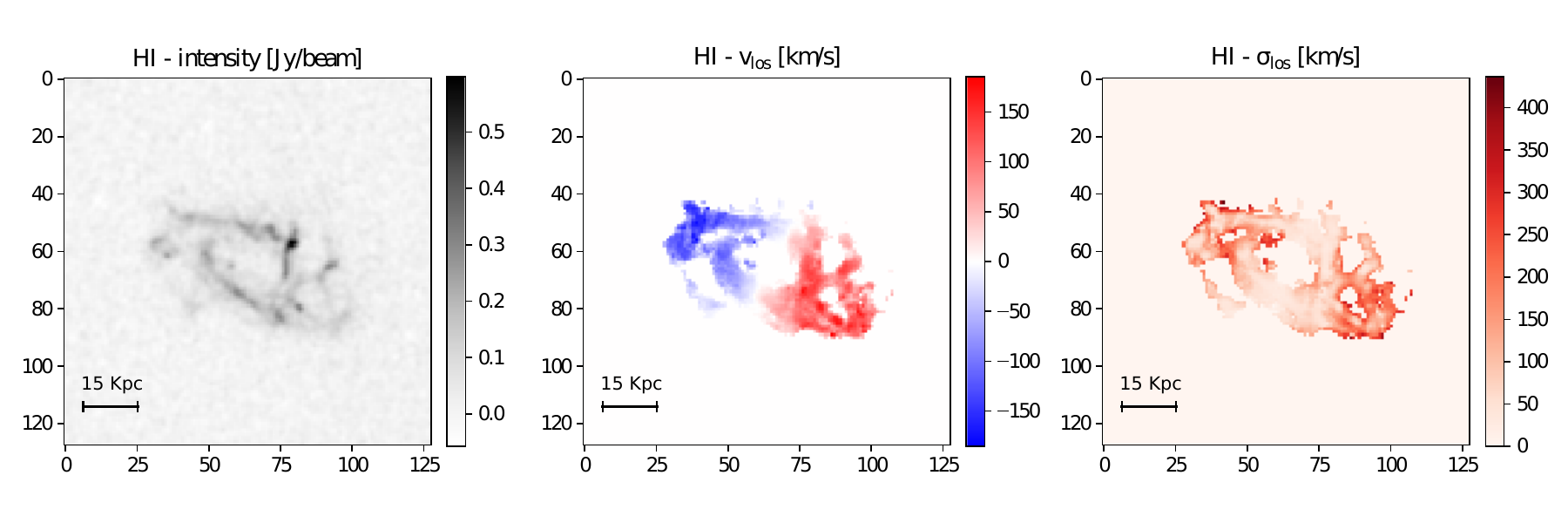}		\includegraphics[width=\linewidth]{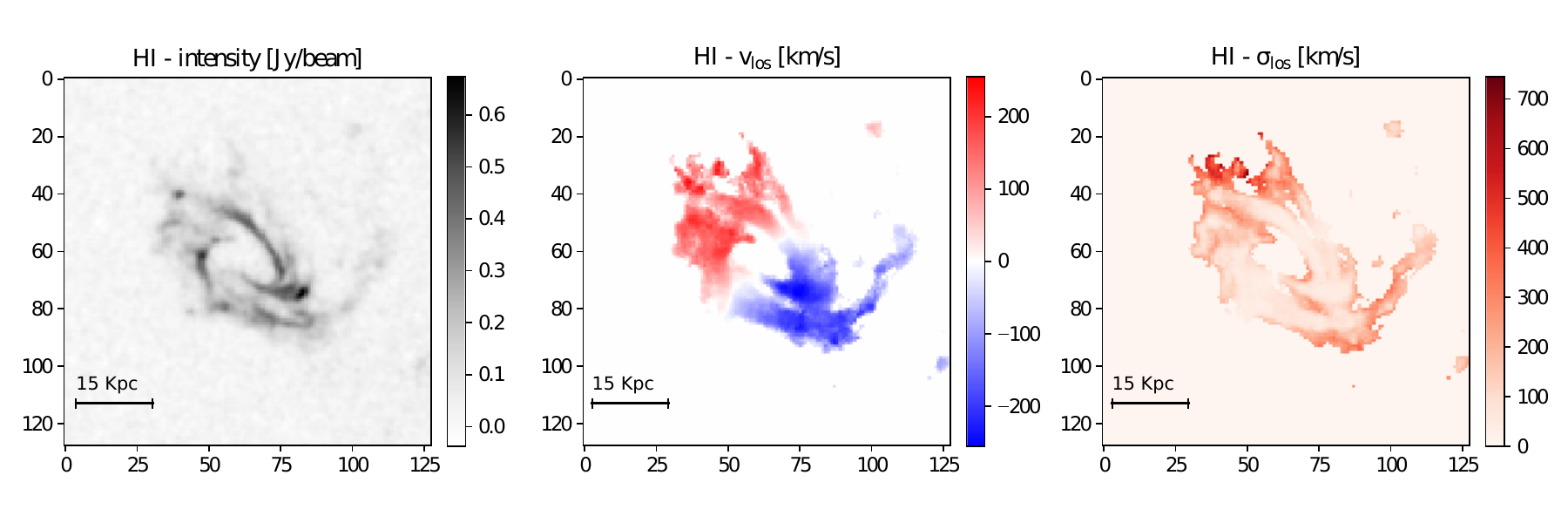}
	\caption{Examples of the mock HI intensity, average velocity and velocity dispersion maps obtained following the procedure described in Sec. \ref{sec:mocks_interferometry} for tow random galaxies (subhaloID = 60744 in the top panel and subhaloID = 108013 in the bottom panel).}
	\label{fig:HI}
\end{figure*}

Using the above setup, the \texttt{MARTINI} code allowed us to obtain realistic mock HI observations. However, since such data cubes consist of 64 individual channels, we decided to reduce the amount of information by projecting out the first three moments of the HI emission. In particular, we computed the corresponding line-of-sight intensity, average velocity and velocity dispersion maps, which can be obtained by evaluating the 0th, 1st and 2nd moments of the emission intensity across the available frequency channels. In order to maintain a good signal-to-noise ratio for the 1st and 2nd moments, we masked regions where the HI intensity dropped below 0.2 Jy/beam (equivalent to an HI column density of $10^{19.5} \, \s{atoms} \, \s{cm}^{-2}$), as suggested in~\cite{Oman2019}. This resulted in the final images, which were later used for training the neural network. In Figure~\ref{fig:HI} we present two examples of synthetic HI maps from our final data set.

\begin{figure}
    \centering
    \includegraphics[scale=0.45]{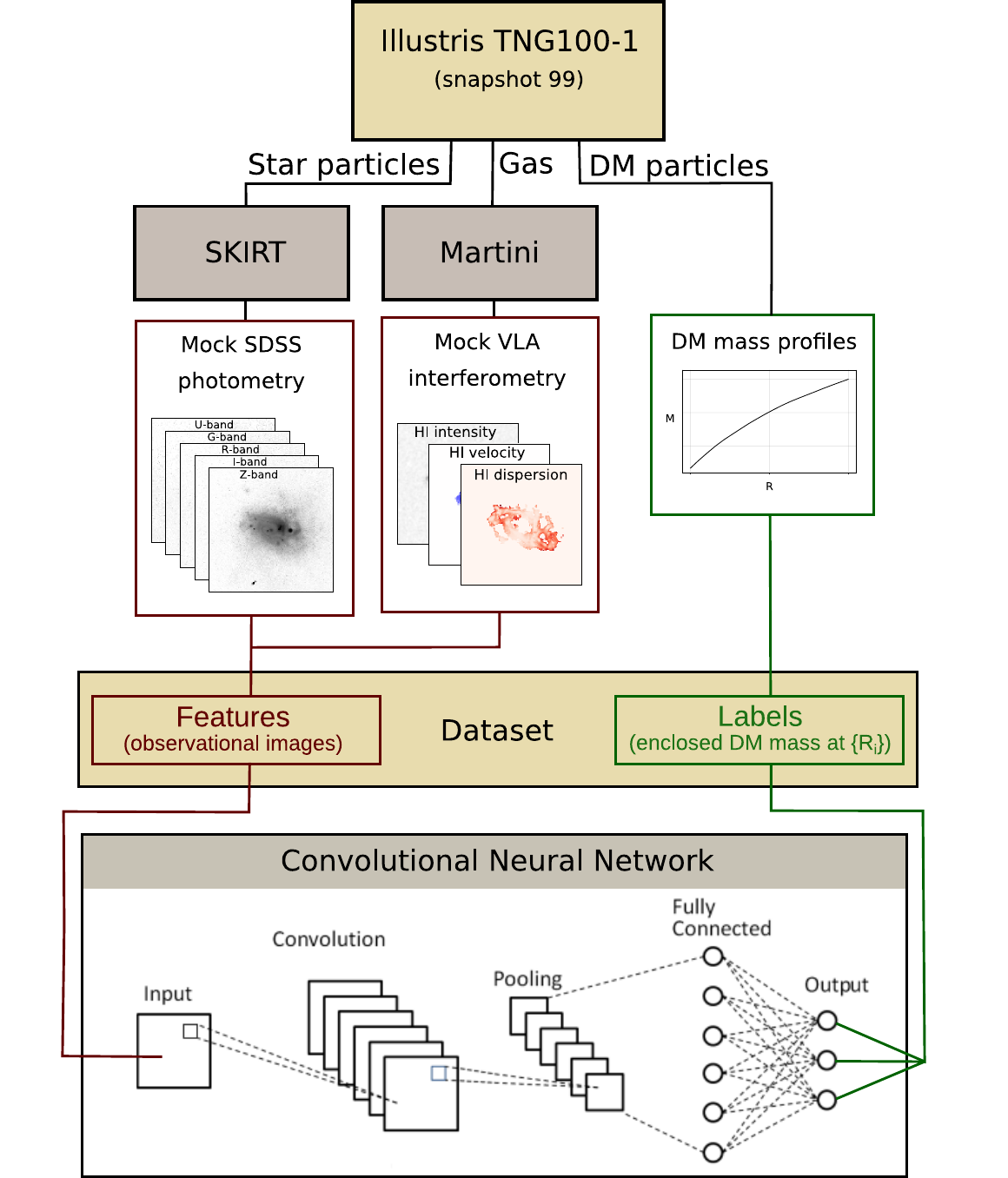}
    \caption{Flow-chart of the analysis pipeline used in our work.}
    \label{fig:flowchart}
\end{figure}

\subsection{Final simulation data sets}
\label{sec:networks_datasets}

Summarising, the benchmark data-set consists of $\sim 6000$ realisations of galaxies for which we create the SDSS photometric images and the corresponding HI data-cube, from which we compute the first three momentum maps.
The photometric images and the three HI momentum maps were stacked to create a 128×128×8 grid (128 by 128 pixel images with 8 channels) that serves as the input to our machine learning model; the output of the network is the DM profile that consist of $20$ scalar quantities, that correspond to the total DM mass enclosed within $20$ radial galactocentric distances that are logarithmically spaced between $1$ and $100 \, \mathrm{kpc}$.
A schematic flow-chart of the pipeline used in this work is depicted in Figure \ref{fig:flowchart}.

As is required in supervised machine learning, we split the data-set into three subsets, containing $60 \%$, $20\%$ and $20\%$ of the total
number of samples that served as training, validation and test sets, respectively.
In order to avoid over-fitting and to have a reliable measurement of the algorithm performance, these subsets must be mutually independent, i.e. the observations (galaxies in our case) that are in one subset cannot exist in another subset.

\section{The Machines}
\label{sec:networks}

\par Machine Learning (ML), seen as a set of techniques and algorithms that allows for robust statistical inferences on data, has proven very useful in physics problems where there is limited knowledge of the physical system (e.g. in the representative case of Indirect Detection searches for DM in astrophysics). 
For a deeper review about machine learning techniques we recommend \cite{Mitchell, Murphy, deepLearning}.
The case at hand is similar: the physical modelling of stellar dynamics in galaxies in the present work is limited and typically relies on simplifying assumptions, which may not always be adequate.  

In this work we deal with a complex data-set describing the kinematics, dynamics, and stellar light emission of galaxies, and aim at inferring the DM content influencing the observed characteristics and behaviour. This is a representative problem of `pattern recognition', for which state-of-the-art deep learning models such as Convolutional Neural Networks are, in general, very good at solving. The popularity of neural networks arises from the fact that these models are `universal approximators' (they can approximate any continuous function under quite mild assumptions); 
since our quantity of interest is the enclosed DM mass within a given galactocentric distance, this corresponds to a regression problem. Given a data-set ${\cal D} = \{{\bf x}_i, y_i\}_{i=1}^N$, with multivariate input ${\bf x}$ and scalar output $y$ (in our case, the DM mass), a possible strategy is to first make a proposition of the probability distribution $p(y|{\bf w})$ that governs our variable of interest $y$, and which depends on a number of parameters ${\bf w}$. For example, we can assume, as is typical, that $y$ follows a Gaussian distribution with mean $\mu({\bf x})$. Then, we estimate the mean by a function $f({\bf x, w})$ given by our ML model, a convolutional neural network containing parameters ${\bf w}$, described in detail below.

Concretely, our input variables correspond to the SDSS photometric images and the first three HI momentum maps, i.e $x_{i} = \{ {\rm SDSS}_{i, u}, {\rm SDSS}_{i, g}, {\rm SDSS}_{i, r}, {\rm SDSS}_{i, i}, {\rm SDSS}_{i, z}, {\rm HI}_{i, 1}, {\rm HI}_{i, 2}, {\rm HI}_{i, 3},\}$; the output variables correspond to the DM mass enclosed within $20$ different radii, i.e $y_{i} = \{ M_{i}(r_{k}) \}$ with $k = 1,...,20$. The training is performed by Maximum Likelihood Estimation of the parameters of the network, where we assume that the output follows a Gaussian distribution. In practice, this is equivalent to minimising the \textit{loss function}, being the mean squared error (MSE) defined as:

\begin{equation} \label{eq:loss}
\mathcal{L}(y_{i}, f(x_{i})) = \frac{1}{20}\sum_{k=1}^{20}\left( \hat{\mu}_{i}(R_{k}) - \mu_{i}(R_{k}) \right)^{2} \, ,
\end{equation}
where $R_k$ corresponds to the radial distance associated with $k$-th output neuron of the network, while $\hat{\mu}_j(R_k)$ and $\mu_j(R_k)$ denote the base 10 logarithm of the true and the predicted enclosed DM mass in the units of $M_\odot$. 

\subsection{Convolutional Neural Networks}
\label{sec:networks_cnn}

When dealing with images, standard networks like the ones defined above -- even if, in principle, are able to successfully perform any regression or classification task --  consider the images as {\it tabula rasa} and do not exploit their spatial structure. For example, they treat equally pixels which are close together or far apart, or do not make use of translation and rotation invariance (e.g. the fact that a cat in a photograph is still a cat, regardless of its orientation or location in a photo). Convolutional networks \citep{Krizhevsky}, or Convnets for short,  are especially designed to work with images while exploiting spatial structure. For the task of pattern recognition, they are faster to train than traditional networks, and thus they can deal with more complex data sets while giving a very high performance. 
\newline\newline\noindent
The ConvNets have some important characteristics:
\begin{itemize}
\item  A datum (pixels of an image in this case) is not treated as a row vector in the data-set, but as a squared array of pixels, like the image itself. Similarly, some of the layers in ConvNets are also a set of stacked squared 2D arrays. These layers are called in Appendix \ref{app:network_description} ``2D convolution''. However they are defined via a non-linear transformation of the input, in a similar way as happens with traditional networks. A typical non-linear transformation used nowadays is the so-called Rectified Linear Unit function (ReLU in Appendix \ref{app:network_description}).
\item Not all the pixels of the image affect all the variables (or units) of the first hidden layer. Instead, different (but partially overlapped) small regions of the input are connected to different units of the convolutional layer. The size of those small regions have to do with the \textit{kernel size}. The degree of overlapping of those regions determines the \textit{stride} length (see Appendix \ref{app:network_description}). 
\item The parameters connecting the input to the first hidden layer are the same (are shared) for all the units. The same applies to the parameters connecting the first hidden layer to the second, and so on. This configuration, together with the previous characteristic, reduces greatly the amount of parameters with respect to the one in a fully-connected architecture. On the other hand it implies that all the units of the hidden layer `learn' the same feature (e.g. an edge, or another spatial structure) of the image, but placed on different spatial locations; this is the way in which ConvNets implement translation invariance. These shared weights define one kernel, and typically ConvNets use several of them (see in Appendix \ref{app:network_description} how different convolutional layers have different number of kernels). 
\item Another type of transformation is \textit{pooling}, which defines a type of layer (see Appendix \ref{app:network_description}). This sort of transformation is typically applied right after a convolutional layer, and the idea is to reduce the information, by going from a 2D array of some size (corresponding to the 2D convolution), to another 2D array of a smaller size. For example, in Appendix \ref{app:network_description}, we use a pooling size of 2x2 pixels, meaning that the convolutional layer has been reduced by a half. The effect of pooling is that, after learning the presence of a spatial structure, its exact location is thrown away; there are many possible ways to reduce information in a pooling layer. A typical one, which is also used in our work, is \textit{max-pooling}, which just takes the pixel with the highest value of the weight in the 2x2 pixel subregion.    
\item Finally, ConvNets have a few dense (traditional, fully connected) layers, which are the ones determining the output. The idea is that, once the spatial structures present in the images are learned by the previous convolutional layers, then a traditional network is very good at predicting the output, whether classification or regression. 
\end{itemize}

In order to reduce over-fitting, we used, apart from the training/test splitting of the data commented above, several common techniques that have been proven to prevent the network from \textit{memorising} the training data, but instead, force a \textit{generalisation} of the data. To this end we used \textit{data-augmentation} (creating different observations of the same galaxy by randomly varying the distance and orientation) and the regularisation technique called \textit{dropout}, that simply discards, at random, some connections inside a hidden layer (being convolutional or dense). As we show in Appendix \ref{app:network_description}, we have used dropout where 25\% of the connections are set to zero.

For all the training we use the Adam optimiser, which is a state-of-the-art implementation of the Stochastic Gradient Descent technique to optimise a function. We use a starting learning rate of $10^{-4}$.

It is worth remarking that the architecture of the network can be arbitrarily chosen.
In principle, increasing the number of layers in a CNN
may improve the results, but at some point, very deep models
will become too difficult to train and may suffer for some instabilities on their performance. 

For this reason we decided to train and test different CNN architectures with an increasing
level of complexity. In Appendix \ref{app:network_description} we described the CNNs used in our analysis.

We also train and test Residual neural networks (ResNets) \citep{resnet}.
This kind of model is a very deep CNN that overcomes some of the problems pointed out above using the \textit{skip connections} method, to reduce the difficulties in training deep CNNs. This is a state-of-the-art architecture that has been used with much success in several astronomical
applications, including detecting strong gravitational lenses \citep{Lanusse2018}, inferring DM subhalos in strong lensing images \citep{dmSubs}, and the identification of Sunyaev-Zel'dovich galaxy clusters \citep{SZClusters}, among many others.

\section{Results}
\label{sec:results}

\par The key goal of this work is to explore the capabilities of convolutional neural networks in reconstructing the DM mass profile of simulated galaxies using different sets of realistic mock observations. Here we present our results, beginning with tests of different neural network typologies, namely three ``custom built" models, as well as the ResNet50, which were described in the previous section. Subsequently, we turn our attention to the impact of using different sets of observations, namely various bands of the SDSS photometric images and projected moments of the HI interferometric data, as well as their different combinations.
\par This provides us with intriguing results, which show that the convolutional neural networks are able to extrapolate the DM mass profile of galaxies with reasonable accuracy directly from the photometric images, avoiding the need for observationally costly kinematic measurements, which are the centerpiece of traditional methods. Following these initial benchmarks, we additionally explore the uncertainties in the predictions of the networks using a method based on bootstrapping; by applying it to a sample of test objects, we demonstrate that it indeed provides us with a reasonable estimate for the errors. We further validate the performance of our network by examining the sensitivity maps, which clearly show that network's predictions are indeed driven by physically meaningful segments of the input images. 

\subsection{Accuracy of the networks}
\label{sec:results_accuracy}

After training the neural networks, we tested their performance by applying them to our test dataset. To quantify their accuracy, we compared their predictions for the enclosed DM mass with the true values using the normalised root mean squared error, $\Delta_\s{MSE}$, and mean absolute error, $\Delta_\s{MAE}$, summary statistics:
\begin{align}
	\Delta_\s{MSE}(R_i) & = \left[ \frac{1}{N} \sum_{j=1}^{N} \left( \mu_j(R_i) - \hat{\mu}_j(R_i) \right)^2 \right]^{1/2} \, , \\
	\Delta_\s{MAE}(R_i) & = \frac{1}{N} \sum_{i=j}^{N} |\mu_j(R_i) - \hat{\mu}_j(R_i)| \, ,
	\label{eq:performance}
\end{align}
where $R_i$ corresponds to the radial distance associated with $i$th output neuron of the network, $j$ runs over all the galaxies in the test set, while $\mu_j(R_i)$ and $\hat{\mu}_j(R_i)$ denote the base 10 logarithm of the predicted and true enclosed DM mass in the units of $M_\odot$.

We note here that Eqs. \ref{eq:performance} are a customary estimate of the performance of the ML output, applied it to $\mu$ and $\hat\mu$. 
In our case, the machines being trained on the logarithm of the masses and being this quantity the output, we used the logarithm of the masses to assess the performance of the network.

\begin{figure*}
	\centering
	\includegraphics[width=\linewidth]{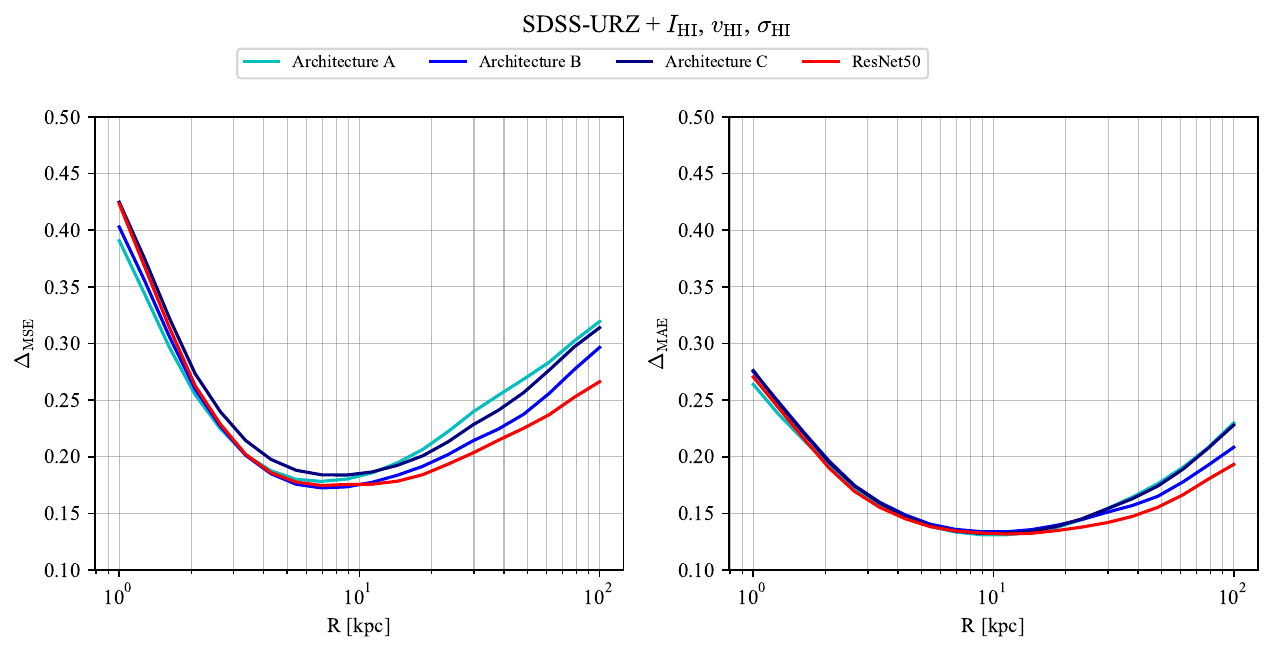}
	\caption{Root mean squared error, $\Delta_\s{MSE}$, and mean absolute error, $\Delta_\s{MAE}$, shown in the left- and right-hand side plot, respectively, as obtained for network architectures A, B, C and ResNet50 when using the \textit{u}, \textit{r} and \textit{z}-band photometric images along with the HI intensity, l.o.s. velocity and l.o.s. velocity dispersion map as inputs for the networks.}
	\label{fig:arch_compare}
\end{figure*}

In Figure~\ref{fig:arch_compare} we show the obtained values of $\Delta_\s{MSE}(R)$ and $\Delta_\s{MAE}(R)$ for the four network architectures considered in this work, trained on 6 observational channels, namely the \textit{u}, \textit{r}, and \textit{z}-band SDSS images and the HI intensity average velocity and velocity dispersion maps (analogous results for different combinations of input data are shown in the Appendix~\ref{app:architecture_comparison}). As can be seen from the plots, all four networks perform roughly equally well: they reach $\Delta_\s{MSE} \lesssim 0.2$ ($\Delta_\s{MAE} \lesssim 0.15$) around $R \sim 10$ kpc, while the errors gradually increase to $\Delta_\s{MSE} \sim 0.3$ ($\Delta_\s{MAE} \sim 0.2$) in the outskirts and rapidly grow at $R \lesssim 3$ kpc, where they can surge up to $\Delta_\s{MSE} \gtrsim 0.4$ ($\Delta_\s{MAE} \gtrsim 0.25$). These trends are in good agreement with our expectations, since the observations provide the strongest handle on the DM content at intermediate radii, where the DM halo begins to dominate the dynamics of the galaxy, but one still has a sufficient amount of baryons to trace its gravitational influence. On the other hand, at the largest few radial points the networks are extrapolating the DM mass profiles well beyond the extent of baryons, which naturally leads to increasing errors.
In the innermost parts (i.e. $R \lesssim 3$ kpc) the performance of the networks rapidly decreases, which implies weak correlations between the distribution and kinematics of visible matter with the underlying DM mass. This is again in agreement with our expectations, since the central parts of the galaxies are known to be dominated by baryonic physics. Furthermore, due to the difficulty of performing accurate measurements, as well as the complexity of the central regions, the traditional techniques of inferring the DM distribution are known to be even less successful in the inner few kpcs of the galaxies (see, e.g., \cite{Oman2019})

Another important result that can be drawn from Figure~\ref{fig:arch_compare} (as well as the analogous plots for different combinations of input data, presented in Appendix~\ref{app:architecture_comparison}), is that all of our benchmark network architectures show similar performance at all radii, despite drastic differences in the corresponding number of free parameters -- as described in the previous section, ResNet50 contains approximately 40$\times$ more free parameters than network architecture A. While there is a slight trend of more complex networks reaching better accuracies, which becomes most noticeable at large radii, the differences are smaller than the typical variations in the performance encountered in retraining the networks from a random initialization state. For this reason, we will in the following adopt network architecture B as our benchmark example; however, we explicitly checked that analogous results hold also for the other three architectures.

\subsection{Comparison of different observational inputs}
\label{sec:results_inputs}

All the benchmark network architectures presented in the previous section lead to good performance in inferring the DM mass profile from a combination of SDSS photometric images and HI interferometric data. However, an equally important question is how the performance varies when only a subset of the aforementioned observations is available.

\begin{figure*}
	\centering
	\includegraphics[width=\linewidth]{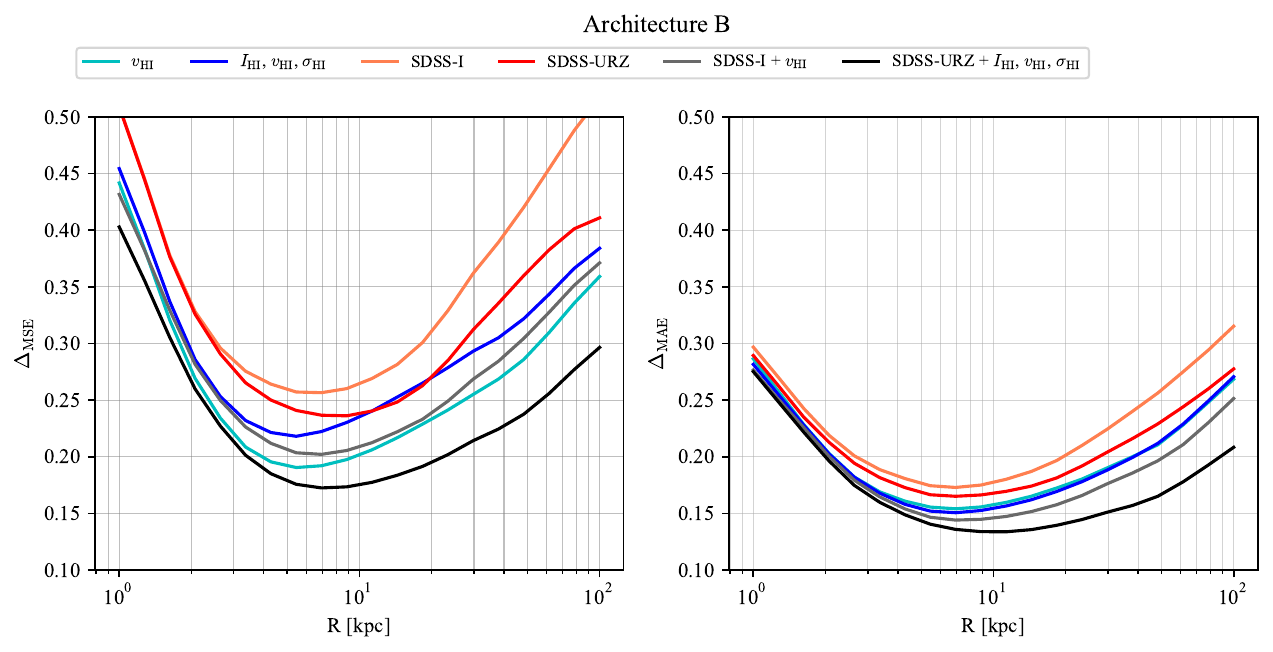}
	\caption{Root mean squared error, $\Delta_\s{MSE}$, and mean absolute error, $\Delta_\s{MAE}$, shown in the left- and right-hand side plot, respectively, as obtained for network architectures B for various combinations of input observations.}
	\label{fig:input_compare}
\end{figure*}

In Figure~\ref{fig:input_compare} we show the obtained $\Delta_\s{MSE}$ and $\Delta_\s{MAE}$ of the network architecture B when trained with various combinations of the input data (analogous results for other network architectures can be found in Appendix~\ref{app:dataset_comparison}). As expected, the best performance is achieved when the network is supplied with the most information, i.e. for the combination of \textit{u}, \textit{r}, and \textit{z}-band photometric images and HI intensity, average velocity and velocity dispersion maps. The errors increase if the networks are provided only with SDSS \textit{i}-band photometric and HI l.o.s. velocity map or HI interferometric data alone. However, the decrease in the performance is relatively minor and most pronounced at large radii, while at intermediate $R \sim 7$ kpc the network still manages to reach $\Delta_\s{MSE} \sim 0.2$ ($\Delta_\s{MAE} \sim 0.15$). Further decrease in performance can be observed if the network is provided only with photometric images. The errors in the central part of the objects increase by roughly the same amount, regardless if only single \textit{i}-band or \textit{u}, \textit{r}, and \textit{z}-band photometry is used. On the other hand, at $R \gtrsim 5$ kpc the three-channel network performs notability better, reaching the minimal $\Delta_\s{MSE} \sim 0.23$ ($\Delta_\s{MAE} \sim 0.17$) around $R \sim 10$ kpc.

A particularly important conclusion, which follows from the above described results, is that our method of inferring the DM mass profiles of galaxies works reasonably well, even if only photometric observations are available. This interesting and unforeseen result appears to be in line with that of similar studies to ours in the literature, particularly \cite{WuBoada}. These authors train a CNN directly on real SDSS imaging and spectroscopic data to find that their ML model learns a representation of the galaxy gas-phase metallicity from the optical imaging alone, even beyond what is normally only accessible through the conventional spectroscopic analysis of oxygen spectral lines. Such successful examples of ML applications in astrophysics lend us the hope and support that, what we are detecting in the present study, is a true correlation between galaxy morphology, resolved features, luminosity (stellar mass by proxy), and the galaxy halo mass. A recovery of the stellar-to-halo mass relation would not be surprising in this case, as galaxy stellar mass is directly measured from image photometry (via a thoughtfully chosen mass-to-light ratio), and has been shown to be tightly coupled to galaxy halo mass (see, e.g., \cite{Girelli2020} and \cite{Posti2020}) and internal structure (see, e.g., \cite{Kauffmann2003}).

Since obtaining resolved HI data cubes is highly time demanding, computationally expensive and currently possible only for several hundreds of galaxies within the local Universe, the possibility of inferring the DM mass profiles directly from photometric images provides a major advantage. Furthermore, the traditional methods of studying DM distribution within galaxies (with the exception of gravitational lensing) all require accurate kinematic measurements, which are equally difficult to obtain. Therefore, our approach appears to provide a unique tool for determining the DM mass directly from vast photometric catalogues which are readily available, such as the one of SDSS. 

\subsection{Error estimates via bootstrapping}
\label{sec:results_bootstrap}

As demonstrated above, the neural networks are capable of inferring the enclosed DM mass with good accuracy over a wide range of galactocentric distances. However, for practical applications it is equally important to provide reliable error estimates for the network's predictions. To achieve this, we adopt a technique based on bootstrapping, which has been suggested as a possible way of providing the error estimates in the given context~\citep{cowan}. In this work, we implement it by re-training randomly initialised neural network 100 times, with each run using a distinct bootstrapped data-set that was obtained from resampling the original training set with replacements (i.e. same object can appear multiple times within each bootstrapped data-set). After all the networks completed the training, one can use them to obtain a distribution of the predictions, for which the central values and the corresponding credibility intervals can be computed. As we demonstrate below, this approach indeed provides us with reasonable error estimates for the predictions of the neural networks.

\begin{figure}
	\centering
	\includegraphics[width=\linewidth]{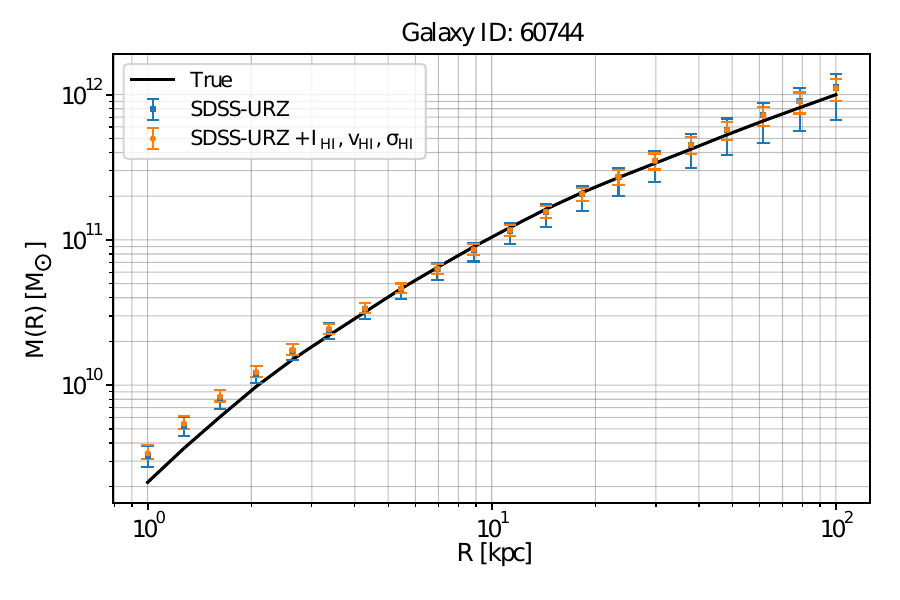}
	\includegraphics[width=\linewidth]{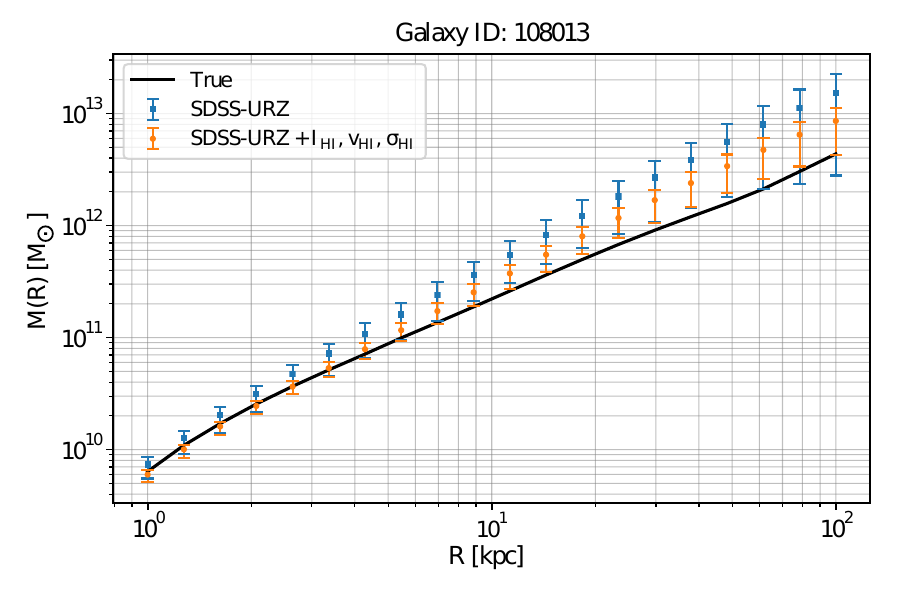}
	\caption{Comparison of the true DM mass profiles with the predictions of neural networks with architecture B for two random galaxies of the test dataset. We show the results for networks trained either on SDSS \textit{u}, \textit{r} and \textit{z}-bands images only (blue), as well as the networks which were additionally provided with HI intensity, average velocity and velocity dispersion maps along the line-of-sight (orange). The central marks correspond the median value while the error bars denote the corresponding 68\% credibility intervals, as obtained through the bootstrapping method.}
	\label{fig:bootsrap}
\end{figure}

In Figure~\ref{fig:bootsrap} we show the median predictions and the corresponding 68\% credibility intervals for neural networks with architecture B, along with the true DM mass profiles for two galaxies from the test set, which were not exposed to the networks during the training procedure. As can be seen from the plots, our networks provide fairly accurate predictions for the DM mass profiles and the true values mostly lie within the error estimates provided by the bootstrapping technique. Furthermore, we explicitly checked that the average size of the 68\% credibility interval predicted by the bootstrapping method provides a good match with the typical value of $\Delta_\s{MSE}(R_i)$ in all the radial bins. In the case of object 108013, shown in the lower panel of Figure~\ref{fig:bootsrap}, we can also see that the bootstrapping technique correctly accounts for the highly asymmetric errors in the predictions. This illustrates that providing the uncertainty estimates on the inferred DM mass profiles is crucial, since using only the median values can induce significant bias due to the possible skew in the distribution of the predictions.

From the same Figure~\ref{fig:bootsrap}, we can additionally appreciate the differences in the performance of the networks that were trained with and without the HI interferometric data. In particular, the networks that were trained using both, photometric and interferometric data, perform noticeably better and also lead to smaller error bars. However, even in the case when only the photometric images were used, the inferred DM mass profiles are reasonably accurate and in most cases bracket the values predicted by the networks that were additionally provided with the HI maps.

As a final note, we stress that the errors obtained through the bootstrapping technique are capable of accounting only for the uncertainties stemming from the non-optimal regression of the neural networks. For practical applications, one would have to additionally account for systematic errors arising from the inaccuracy of hydrodynamical simulations themselves. Unfortunately, assessing the latter is still a topic of active research and falls beyond the scope of this work.

\subsection{Understanding the neural network output with sensitivity maps}
\label{sec:results_sensitivity}

\begin{figure*}
	\centering
	\includegraphics[width=0.75\textwidth]{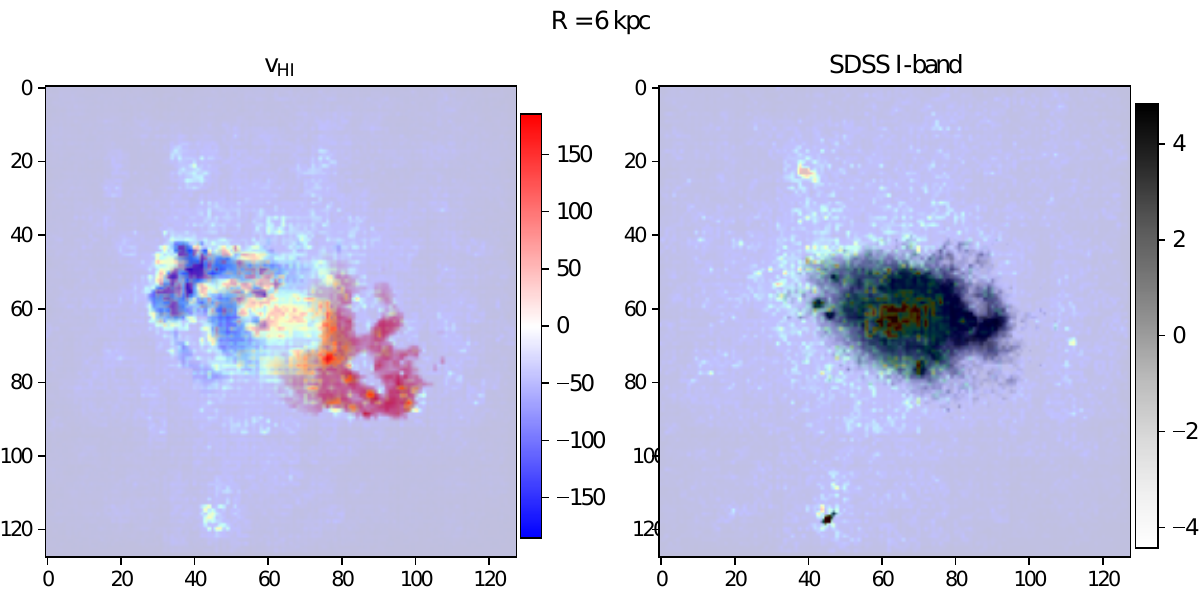}
	\includegraphics[width=0.75\textwidth]{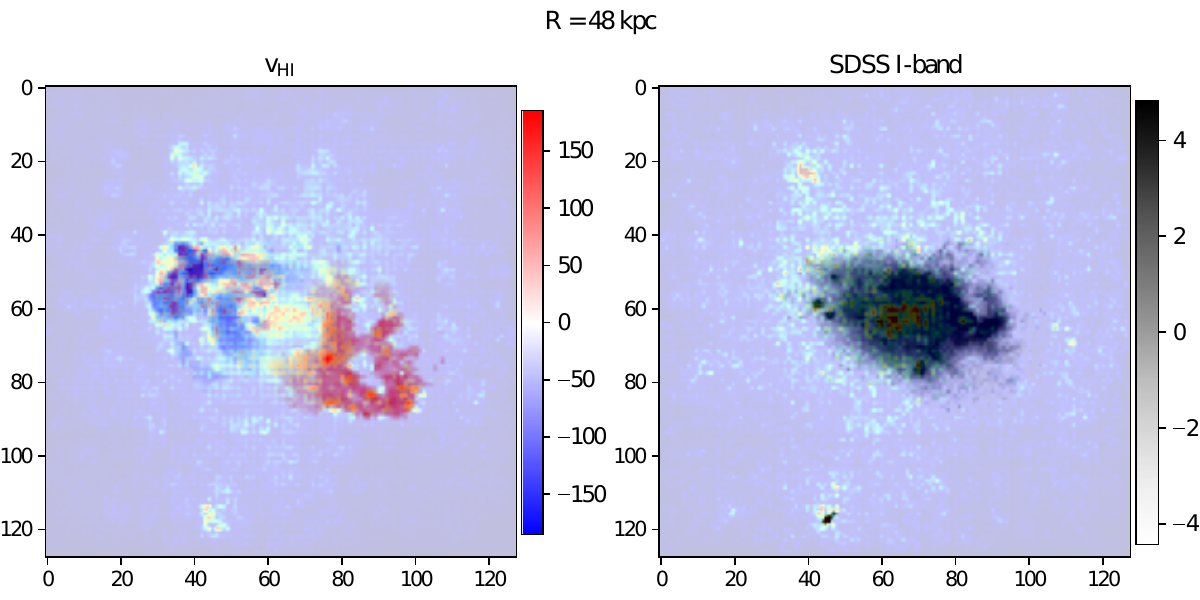}
	\caption{Input images overlaid with the corresponding sensitivity maps for neural network with architecture B that was trained using SDSS \textit{I}-band photometry and HI average velocity along the line-of-sight. The colour-coding and units for $v_\mathrm{\small HI}$ and SDSS I-band maps are same as in figures~\ref{fig:HI} and~\ref{fig:SDSS}, respectively, while green colouring is proportional to the pixel importance computed using the \texttt{SmoothGrad} algorithm (we omit the corresponding scale since we are primarily interested in a qualitative properties).}
	\label{fig:sensitivity_maps}
\end{figure*}

While the traditional methods for inferring the DM content of a galaxy are based on well established physical laws, the inner workings of neural networks are, to a large degree, opaque. This often represents a crucial weakness of studies based on machine learning, since performing consistency checks in the analysis pipeline is notoriously difficult, if not impossible. This issue opens up a whole new area in the machine learning community, that searches for methods that allow us to have a deeper understanding and interpretation of the ML models. For a deeper review on methods that explain machine learning results, we recommend \cite{gilpin2018}.

In the case of CNN, several methods have been proposed that offer validation that the networks are indeed relying on the input features which are known to be crucial for the given task. In this work we adopt the approach of inspecting the sensitivity maps (sometimes also referred to as \textit{saliency} maps), which allow us to quantify the importance of each pixel in the input images for obtaining a given result. Formally, the sensitivity map can be defined as the derivative,
\begin{align}
	\label{eqn:sensitivity_map}
	S_{ij} \equiv \frac{\partial y}{\partial x_{ij}} \, ,
\end{align}
where $y$ denotes the output of the network, and $x_{ij}$ represents the pixel corresponding to coordinate pair $(i, j)$ in the input image.
In the case of multiple input channels and output variables, we will have a distinct map for the sensitivity of each channel in predicting each output neuron. 
On the other hand, in practice it turns out that with the above definition, sensitivity maps are often very noisy and can be very different if a tiny change is introduced in the inputs.
To address this issue, we follow the so-called \texttt{SmoothGrad} approach, introduced in~\cite{smoothgrad}. We used the numerical implementation that is publicly available in the \texttt{tf-keras-vis} module~\citep{keras_vis}, which advocates for averaging the sensitivity maps obtained via Equation~\eqref{eqn:sensitivity_map} over a sufficiently large set of slightly perturbed input images. 
This allows us to obtain smoother sensitivity maps and, hence, establish more reliably which parts of the input image are most important for the CNN.

In Figure~\ref{fig:sensitivity_maps} we show the input images overlaid with the smoothed sensitivity maps of the mass enclosed within $6$ kpc and $48$ kpc (top and lower panels respectively) for the benchmark neural network (architecture B) trained on SDSS \textit{i}-band photometry and HI line-of-sight velocity information. The color-coding and units for the input images are the same as in Figures~\ref{fig:HI} and~\ref{fig:SDSS}, respectively, while the sensitivity map is displayed in green scale proportional to the pixel importance computed using the \texttt{SmoothGrad} algorithm. Hence, white pixels represents regions with very low importance on the final predictions, while, on the other hand, darker pixels represents regions with higher importance on the final output.
As can be seen from the plots, the gradients of the outputs (and hence the sensitivity of the output to those pixels) are the largest when taken with respect to the central part of the input images, as one would naively expect. 
Furthermore, the sensitivity maps seem to nicely follow the irregular shapes of the input images, while the blank areas (i.e. parts of the input images that are devoid of baryonic emissions) bear no significance for the final result. Additional confirmation that the networks work as expected comes from comparing the top and bottom panels of the same Figure~\ref{fig:sensitivity_maps}, which show the sensitivity maps for the output neurons corresponding to the enclosed DM mass at $R = 6$ kpc and $R = 48$ kpc. While the general patterns are similar in both cases, the sensitivity maps obtained for the outer radius is noticeably more sensitive to the outskirts of the central galaxy, as well as the small satellite galaxies that are visible in the photometric image. These considerations provide us with compelling evidence that the neural network is indeed using the physically relevant features in the input images to infer the DM mass profile.

\subsection{Comparison with Tully-Fisher relation}
\label{sec:TF_comparison}

\par In this section we perform a comparison between the results we obtain through the use of the ML algorithms developed and described until now and those obtained by a Tully-Fisher relation. The goal of this section is to understand whether the uncertainties in the outcome of the ML method here proposed will make the method competitive wrt existing empirical relations, such as the TF, once it will be fine tuned to be applied to a real-case scenario.
With that in mind, within the synthetic Universe of the simulations only, and keeping also in mind that the physical consistency of the simulation (i.e.: the capacity of the simulation to reproduce empirically known relationships) has been tested by their developers \citep{vogelsberger_2014_2}, we have chosen one formulation of the empirical Tully-Fisher relations, and compared the uncertainties introduced by its formulation with those introduced by our method.

The Tully-Fisher \citep{tully-fisher} relation is an empirical relationship  that correlates the luminosity  of a spiral galaxy with its rotation velocity. 
It provides a way to estimate the total mass of a spiral galaxy based on its observable properties.
In \cite{TF} they have calibrated the relation using $3041$ spiral galaxies from the SDSS DR7 with available rotational velocities from HI line widths, and found the best-fit parameters:

\begin{equation}
    \rm Log (V_{\rm rot} ) = (0.264 \pm 0.010) \rm Log( M_{\star} ) - (0.558 \pm 0.101)
    \label{eq:TF}
\end{equation}

where $V_{\rm rot}$ is the circular velocity at twice the stellar half-mass radius $r_{\star}$, and $M_{\star}$ is the galaxy stellar mass.

In a real-Universe scenario, one uses the observed stellar mass from photometric images, and by making use of the above relationship,  makes prediction of the expected $V_{\rm rot}$, which is a proxy for the total gravitational mass enclosed in the mentioned radius.

Here, we work in a synthetic Universe where our goal is the following: to
compare the intrinsic spread introduced by the physical relationship in Eq. \ref{eq:TF} with that introduced by the ML method.

To this aim, we select a group of galaxies within the simulation. 
First, for each galaxy within the sample we assess, in turn, the {\tt real} total mass, the rotational velocity, and the {\tt real} baryonic mass, all known from within the simulation. 
Then, for each galaxy we adopt two separate and different procedures to obtain the following quantity, a proxy for the total mass of the Galaxy, $V_{\rm rot}$($r_{\star}$). In order to estimate it using the two methods at comparison here, for each galaxy within the sample we adopt:

\begin{itemize}
    \item Procedure A: Estimate the rotational velocity at twice the stellar half-mass radius using the DM mass predicted by our ML method, $V_{\rm rot}=\sqrt{\frac{G \left[ M_{ML}(R) + M_{b}(R) \right]}{R}}$, to which we add the real baryonic mass estimated from the simulation. This procedures relies on the use of our ML algorithms developed and described in this paper.

\item Procedure B: Assess the rotational velocity at twice the stellar half-mass radius with the empirical Tully Fisher relationship in Eq. \ref{eq:TF}, by using as input the stellar mass as known from the simulation (in fact, we could use the estimate provided by photometric mock images, but they introduce an additional uncertainty which we prefer to ignore here for the goal of this test).
This procedures relies on the use of our the TF as a predictive empirical relation.
\end{itemize}

\begin{figure}
    \centering
    \includegraphics[width=0.5\textwidth]{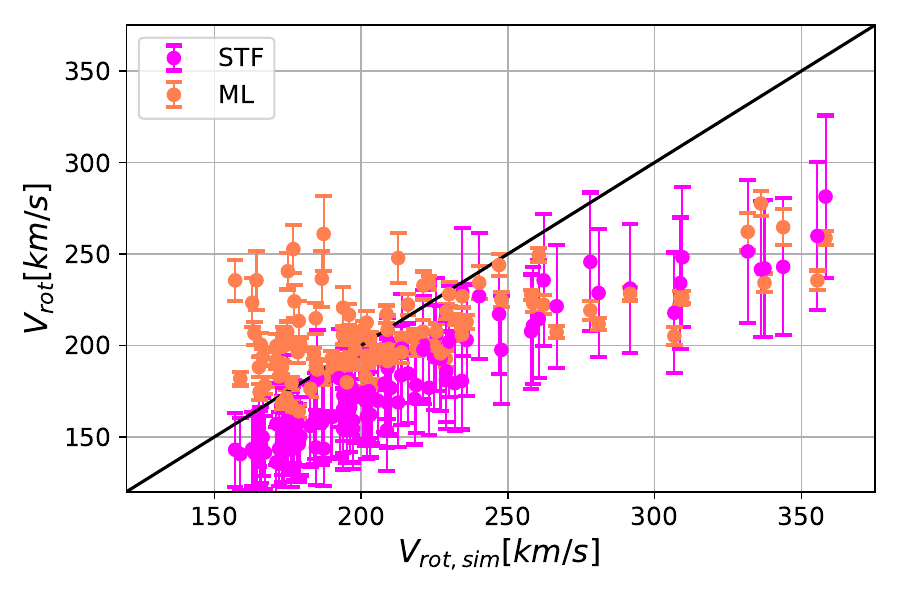}
    \caption{Comparison between the rotational velocity estimated with the machine learning vs the one obtained by using a Tully-Fisher relation.}
    \label{fig:TF-ML}
\end{figure}

In Figure \ref{fig:TF-ML} we show the results of both procedures --{\bf A} in orange, {\bf B} in magenta-- placing the {\tt actual} rotation velocity $V_{\rm rot} (r_{\star})$, as known from within the simulation, on the X axis.

While keeping in mind --as mentioned above-- that the Tully Fisher relationship has been validated in the simulation Universe by the team developing the latter (see Fig. 23 of \cite{vogelsberger_2014_2}), here we only perform the comparison between the outcome of Procedures \textbf{A} \& \textbf{B} with the goal of understanding whether the uncertainties in the outcome of the ML method here proposed will make the method competitive once it will be fine tuned to be applied to a real-case scenario.

From our test, we learn that:
a) ML and TF results are definitely compatible within uncertainties;
b) the uncertainties of the ML method are smaller both individually (statistics), and considering the spread as a systematic envelope; 
c) the ML method tends to slightly overestimate the total mass at the low mass end of the Galaxy sample considered, whereas the TF tends to underestimate it. At the higher mass end, they follow the same trend.
Whereas the latter is certainly a point to be examined in detail in future studies, most interesting in this context is the information one obtains about the uncertainties.
The individual uncertainty (on the total mass /rotation velocity) obtained for each data point / galaxy, is typically smaller in the ML case, than it is from the TF prediction.
Also, the one sigma spread band for the ML case is smaller than the same uncertainty band for the TF, thus making the ML method potentially predictive in a future real Universe application.

\subsection{Comparison with the Rotation Curve method}
\label{sec:results_comparison}

The DM density profile of a disk galaxy is customarily inferred through the analysis of its stellar and/or gaseous rotation curve, as the `missing' component that is required to explain the rotational speeds that appear too large to be explained purely by the visible components of matter. One key aspect of this analysis is the assumption that the disk is rotationally supported and that the chosen physical tracer of the gravitational potential, HI gas in the present work, follows circular orbits, although there are several hints that in some systems this is not the case \citep{Dalcanton}. If some of these assumptions are not fulfilled, unavoidable systematics will be introduced to the final results (see, e.g., \cite{Oman2019}). It is interesting to note that, although this method has been widely used in the literature for many years, there is no clear analysis, to the best of our knowledge, on the precision and accuracy to which this method recovers the true underlying DM profile nor about the systematics introduced.

In this work, we conveniently possess knowledge of the individual components of galaxies in a synthetic Universe, and therefore are empowered to test the Rotation Curve method in an absolute sense, and also in comparison to our ML tool described in the paragraphs above.
While a full description and comparison of the performances obtained with different methods for reconstructing the DM profiles of disk galaxies is beyond the scope of this paper, we present here a handful of examples of synthetic galaxies for which we have performed such analysis, directly comparing the profile of enclosed mass as reconstructed from both the Rotation Curve and ML methods, to the actual profile known from the simulations. To this end, we randomly select four synthetic galaxies to be analysed with both the Rotation Curve and the ML methods developed here.

For each of the simulated galaxies in our test sample, we construct the `observed' rotation curve
by analysing the mock observations of the HI data-cube, using the \texttt{$^{3D}$BAROLO} code \citep{barolo}. It is to be noted that this method is used to generate a realistic rotation curve through an understanding of the natural processes taking place in a physical galaxy: namely via the estimation of the velocity of a chosen target tracer of the rotational speed of the disk at the given position -- in this case, the HI gaseous component.\footnote{This technique is different, and more complex than that adopted in most of the literature, where the total enclosed mass at galactocentric distance is computed from within the enclosed mass as read within the simulation, and a rotation velocity constructed simply for that amount of mass. \citep{marasco_2020}} Through this process we are therefore in possession of the `observed' rotation curve, given the method's assumptions for tracing the total gravitational potential of the galaxy.

Equipped with the mock galaxy rotation curves, we reconstruct the DM distribution in these galaxies through standard methods. First, to obtain the contribution of the visible component, which we take as purely stellar emission, we assume an exponential light profile $L(r)$ to describe the disk, and perform a fit (using the \texttt{AUTOPROF} python software \citep{autoprof}) to the photometric SDSS \textit{g}-band images of the synthetic galaxies. The resulting luminosity profiles are multiplied by an appropriately chosen mass-to-light ratio to obtain a stellar mass profile: we employ a Bayesian analysis (discussed further below) to sample the best-fit mass-to-light ratio for each galaxy from the range of values prescribed by \cite{bell2003} for the measured \textit{L(r)} value and expected range of SDSS \textit{g-r} colours  of the synthetic galaxies. 
For a given mass-to-light ratio $\aries_{\star}$, and assuming that the stars are distributed in a thin exponential disk, we estimate the stellar mass contribution to the rotation curve $v_{\star}$ from the light profile calculated by \texttt{AUTOPROF} (eq. 11.30 of \cite{GalaxyFormation&Evolution})

\begin{equation} \label{eq:stellar_cont}
    v_{\star}^{2} = -4 \pi G \aries_{\star} \Sigma_{0}R_{d}y^{2} \left[ I_{0}(y)K_{0}(y) - I_{1}(y)K_{1}(y) \right]
\end{equation}

where $G$ is the Newton constant,  $\Sigma_{0}$ is the surface density distribution, $R_{d}$ is the scale length of the disk, $I_{0,1}, K_{0,1}$ are the modified Bessel functions and $y=R/R_{d}$.

Second, we model the observed rotation curve as the sum of the DM and stellar contributions \citep{Lelli2016, Li2020},
\begin{equation}
    v_{obs} = \sqrt{v^{2}_{DM} + v^{2}_{\star}}
\end{equation}

where the DM contribution to the rotation curve $v_{DM}$ can be computed analytically assuming an NFW profile for the DM distribution that depends on $V_{200}$ and $C_{200}$ \citep{NFW}.

In addition, as explained in \cite{Li2018}, we apply the following corrections to the measured rotational velocities introduced by uncertainties on the distance ($D$) to and inclination ($i$) of each galaxy:

\begin{equation}
v'_{\star} = v_{\star}\sqrt{\frac{D'}{D}} \quad;\quad v'_{obs} = v_{obs}\frac{sin(i')}{sin(i)}
\end{equation}

Summarising, we have $5$ free parameters that we fit to the observed rotation curve; these correspond to $2$ parameters that describe the DM distribution ($V_{200}$ and $C_{200}$), $1$ parameter that models the mass-to-light ratio ($\aries_{\star}$), and $2$ parameters that describe the geometry of the system ($D$ and $i$).

For the fitting, we perform a Bayesian analysis using the python package \texttt{emcee} \citep{Foreman_Mackey_2013}. We apply a conservative flat prior on $\aries_{\star}$ from $[0-20]$, in order to include the expected values as described by \cite{bell2003}.
In addition, we impose a Gaussian prior on $D$ (around its real value) and on
$i$ (around the value estimated with \texttt{$^{3D}$Barolo}).
For the parameters $V_{200}$ and $C_{200}$ we impose a flat prior around $[10-800]$ and $[1-1000]$, respectively.

\begin{figure*}
	\includegraphics[width=0.4\linewidth]{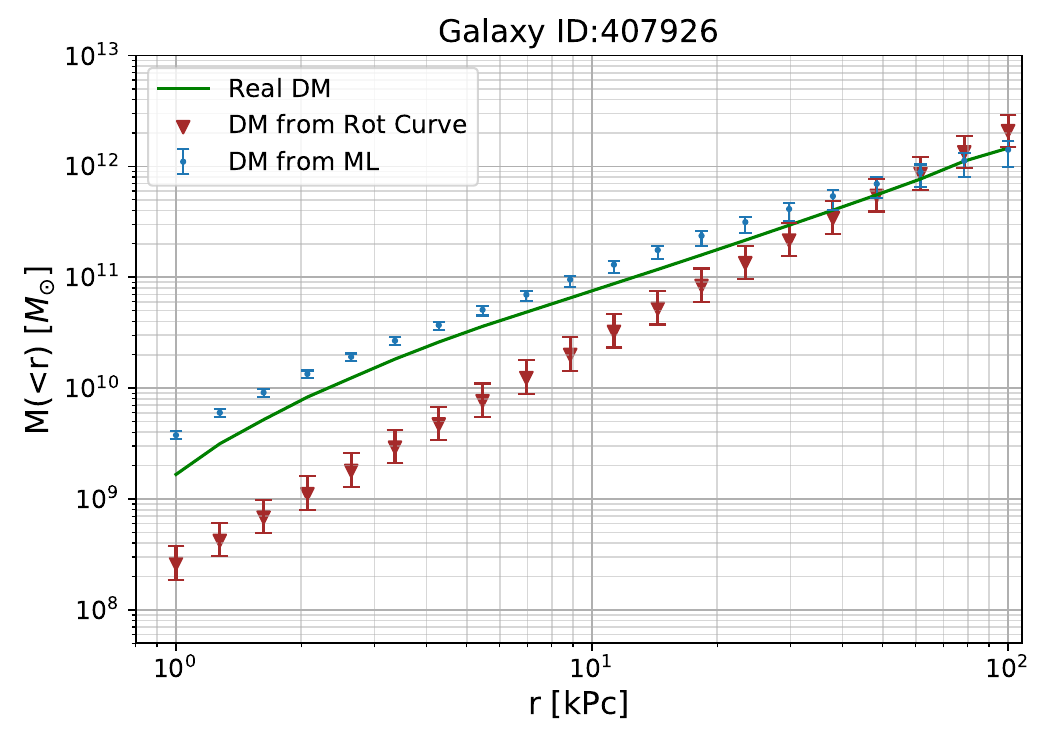}
	\includegraphics[width=0.4\linewidth]{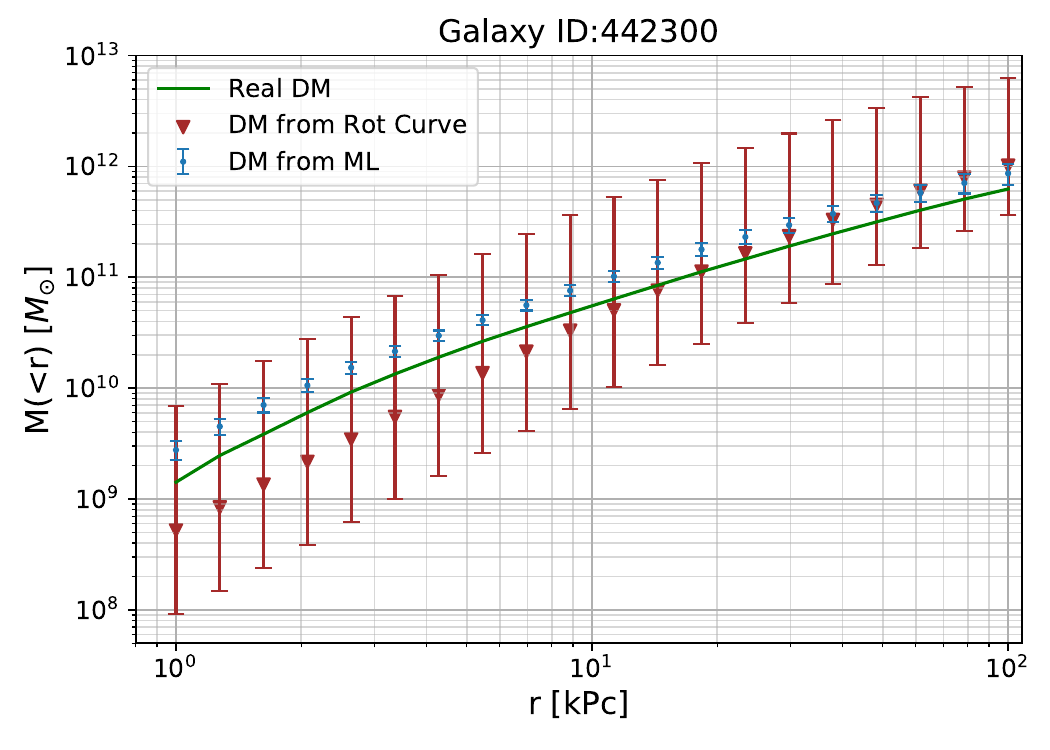}
	\includegraphics[width=0.4\linewidth]{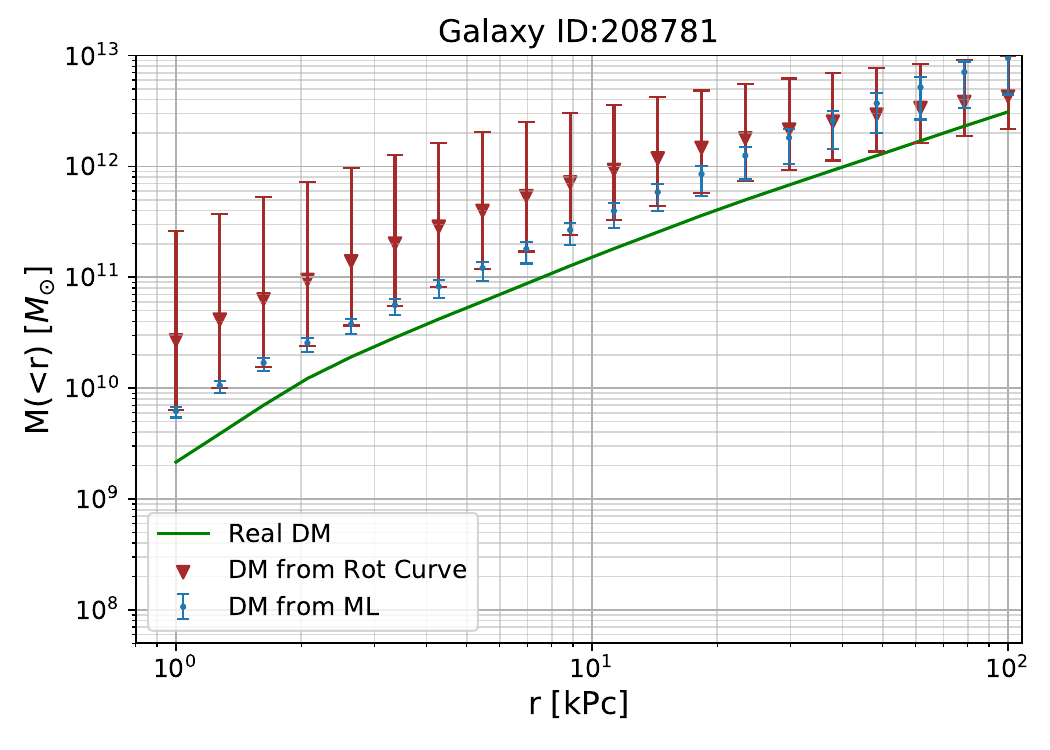}
	\includegraphics[width=0.4\linewidth]{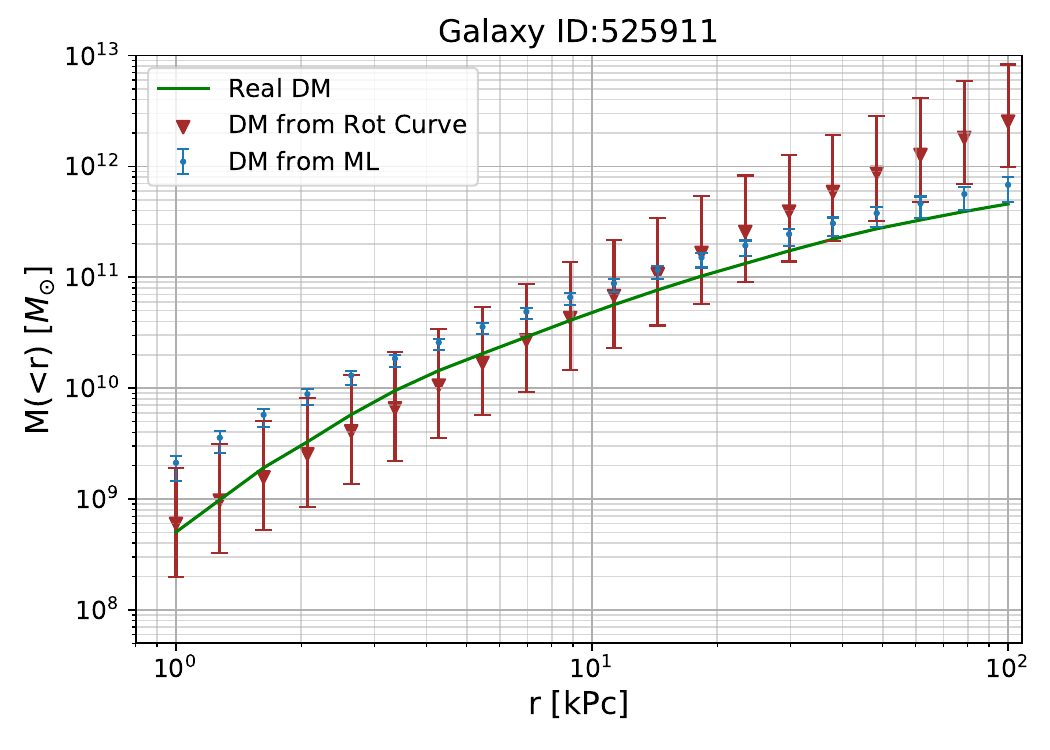}
	\caption{Comparison between the DM profile estimated through different methods and the real one depicted by the green line. The brown triangles trace the result obtained with the rotation curve analysis, while the blue dots represent the results obtained with the ML method.}
	\label{fig:FullComparison}
\end{figure*}

To test the reliability of our Rotation Curve analysis described above, we performed a sanity check that considers the idealised scenario that bypasses the uncertainties arising from this ``observational'' procedure, and instead measures directly the DM distribution already known from the simulated galaxy data, for the $4$ simulated galaxies of Figure \ref{fig:FullComparison}.
For this test, we first extracted the stellar mass profile (i.e. the stellar mass enclosed at different radii) from the $3D$ distribution of stellar simulated particles; and from this profile, we computed the actual stellar contribution to the rotation curve $v_{\star}$. 
thus computed the actual rotation curve (proxy of the total gravitational potential) of the galaxy with the assumption of a rotationally supported disk (crucial to the Rotation Curve method).
We fed the first curve as that of the ``baryonic'' component, and the second as that of the ``observed'' rotation curve to our Bayesian analysis algorithm described above, and reconstructed the Dark Matter density profile by fitting an NFW spherical distribution.
We found that the best-fit NFW profile thus reconstructed agrees very well with the Dark Matter distribution found in the simulated galaxies.
This sanity check proves that our adopted RC procedure is able to recover the physical reality underneath observational data within the controlled environment of the simulations, in an idealized case, without the inevitable uncertainties arising from the ``observational'' procedure.

It is worth noting that in real observations we will not have access to the actual stellar mass distribution nor to the actual rotation curve, and we only may infer those profiles from the photometry and spectroscopy of a galaxy; this, in turn, will introduce uncertainties on these quantities. In order to mimic this behaviour, in the following we assume a $10\%$ uncertainty on both quantities, in line with the median uncertainties in the observed quantities \citep{barolo}.

In Figure \ref{fig:FullComparison}, we display the `real' DM distribution as inferred directly from the output of the TNG100-1 simulation suite (green solid line), and the rotation curve fitting (in brown triangles) that shows a reasonable but unclear pattern of performance. Whereas a systematic analysis of the validity of the rotation curve method within a controlled environment -- and its validity in the real Universe -- is not a goal of this paper, we would like to attract attention to an interesting feature of our rotation curve results: the diversity of their shape, and the non-regularity of their behaviour with respect to the `actual' DM profile from within the synthetic Universe. This could offer an element of reckoning on the persisting problem of ``rotation curve diversity''~\citep{2022NatAs.tmp..130S}.

Compared to the rotation curve method, the ML algorithm developed here provides at face value a better agreement with the input data along a more consistent pattern.
It is worth reminding the reader that the tests provided here apply within the `synthetic' Universe: both the ML learning and the rotation curve tools are used on mock data of star-forming galaxies with masses spanning two o.o.m. around that of the MW, and compared against the DM distribution within that Universe.  The validity of these results when analysing real galaxies will depend on the degree to which the cosmological simulations mirror reality. Also, a careful assessment of reconstruction performance in the presence of instrumental effects will be needed in order to analyse `real' images of galaxies.
Nonetheless, our approach and method of comparison demonstrates the potential ability of machine learning methods to improve upon current traditional analyses.

\section{Conclusions}
\label{sec:conclusion}
\par We have addressed the problem of determining the DM mass distribution in galaxies through a novel approach based on machine learning.

We have made use of the Illustris-TNG100 suite of simulated galaxies in a 
cosmological framework, and have used this as a well-controlled environment to develop, test and train convolutional neural networks. 

We have presented a thorough analysis of the architectures considered, tested with an increasing level of complexity until satisfactory performances were achieved. 

\par Our algorithm is able to reconstruct the DM distribution profile with high performance throughout the full extent of a galaxy, achieving the highest in the intermediate regions with a mean square error below $0.20$ or $0.30$ by using all the available input information, or only the photometric information, respectively.

We stress that our reconstruction of the DM distribution is completely data-driven, and does not require any assumption on the shape or the functional form of the DM profile.

The method developed here is applicable to different types of star-forming galaxies (spiral, elliptical, turbulent galaxies, ongoing mergers) since it does not rely on explicit physical assumptions regarding the dynamical state of the system. \\

We have also shown that convolutional neural networks are capable of reconstructing the DM distribution beyond the radial range covered by baryons, thus allowing a determination of the DM mass in the outskirt of the halo within a unified framework, while traditional methods require extrapolations based on different types of approximations and scaling laws. 
Our ML method also performs better than the traditional Rotation Curve method, reconstructing the real (synthetic) DM distribution more accurately within our synthetic Universe.

The results achieved have been obtained for galaxies with total dark-plus-baryonic masses in the range $\sim 10^{11}-10^{13} M_{\odot}$, but the methodology can be extended to 
a broader mass range. \\

Eventually, training and testing the network over different simulation suites that implement different sub-grid physics models can worsen the performance of the algorithm. However, this bias is expected to be within the standard deviation uncertainty, as recently shown in \cite{Villanueva-Domingo:2021dun}.

The method presented here has been developed on simulated images that were modified to emulate real galaxy images to a high approximation. We anticipate that an application to real galaxies is possible within our framework and that a dedicated applicability study needs to be performed. This will be presented in a forthcoming analysis. 

\section*{Acknowledgements}

We thank the organisers of the ``Advanced Workshop on Accelerating the Search for Dark Matter with Machine Learning''\footnote{http://indico.ictp.it/event/8674/overview}, Trieste, April 2019, where the project was 
conceived as one of the \texttt{darkmachines}\footnote{\url{https://darkmachines.org/}} project challenges.

MdlR acknowledges financial support from the Comunidad Aut\'onoma de Madrid through the grant SI2/PBG/2020-00005 and Funda\c{c}\~{a}o de Amparo \`a Pesquisa do Estado de S\~{a}o Paulo (FAPESP, SP, Brazil) through the process 2019/08852-2.

MP acknowledge partial support from the ANR project ANR-18-CE31-0006 and ARRS grant no. P1-0031. The authors are grateful for the computational resources and the related technical support provided by the IT department of Laboratoire Univers et Particules de Montpellier (LUPM).

BZ has been supported by the Programa Atraccion de Talento de la Comunidad de Madrid under grant n. 2017- T2/TIC-5455, from the Comunidad de Madrid/UAM “Proyecto de Jovenes Investigadores” grant n. SI1/PJI/2019-00294, as well as from Spanish “Proyectos de I+D de Generacion de Conocimiento” via grants PGC2018-096646-A-I00 and PGC2018-095161-B-I00. 
BZ finally acknowledges the support from Generalitat Valenciana through the plan GenT program (CIDEGENT/2020/055).

F.~I. has been partially supported by the Italian grant 2017W4HA7S “NAT-NET: Neutrino and Astroparticle Theory Network” (PRIN 2017) funded by the Italian Ministero dell’Istruzione, dell’Università e della Ricerca (MIUR), and Iniziativa Specifica TAsP of INFN.

\section*{Data Availability}

The data underlying this article will be shared on reasonable request to the corresponding author.

\section*{Note added}
While this paper was in its last stages before completion, the following analysis appeared on the on the arXiv. 
\\
In \cite{buck2021predicting} the authors apply a similar idea, using convolutional neural networks but to infer the properties of the ``visible'' component of galaxies instead.
\\
The idea in \cite{vonMarttens21} is similar to ours, also training on a similar galaxy sample. However, methods and details of architecture analysis are different. Results are compatible.
\\
The analysis \cite{Villanueva-Domingo:2021dun} appeared on the arXiV the very day of our paper. It contains a similar idea, training different architectures on galaxy samples, with different methodologies and scope.



\bibliographystyle{mnras}
\bibliography{references} 




\appendix

\section{Dataset comparison (for a fixed network)}
\label{app:dataset_comparison}

\begin{figure*}
    \centering
    \includegraphics[width=\linewidth]{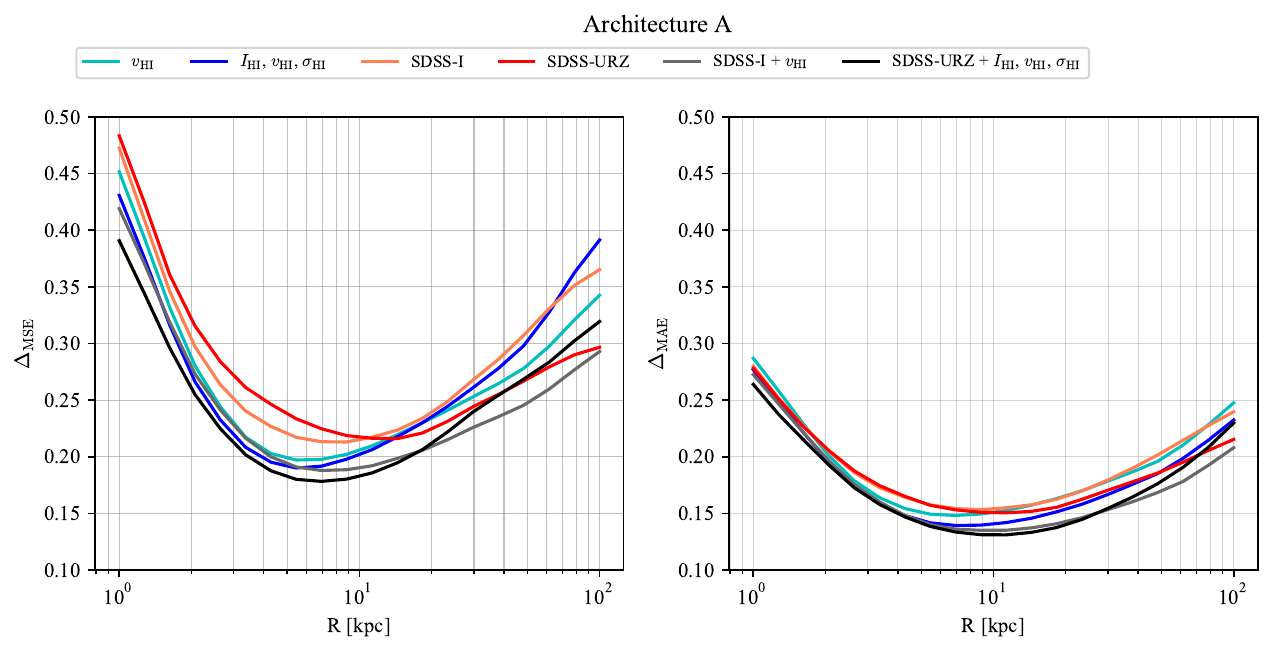}
    \caption{Root mean squared error (left) and mean absolute error (right) obtained for different sets of observations in the case of neural network architecture A}
    \label{fig:results_modelA}
\end{figure*}

\begin{figure*}
    \centering
    \includegraphics[width=\linewidth]{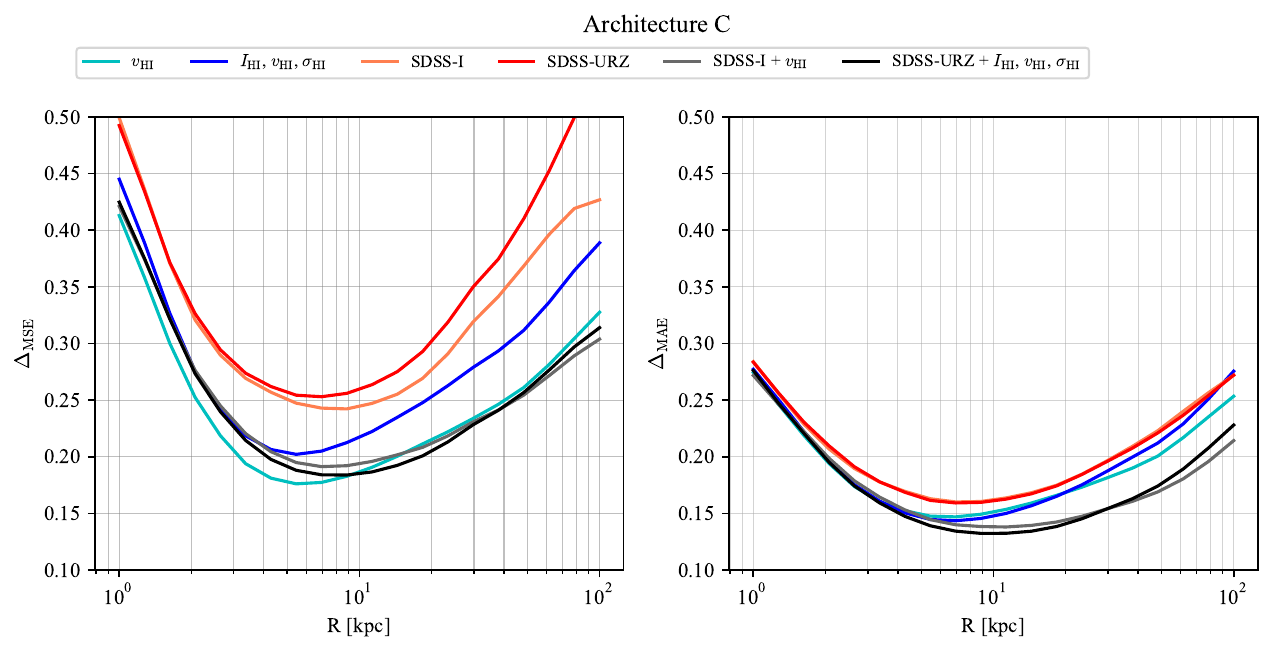}
    \caption{Root mean squared error (left) and mean absolute error (right) obtained for different sets of observations in the case of neural network architecture C}
    \label{fig:results_modelC}
\end{figure*}

\begin{figure*}
    \centering
    \includegraphics[width=\linewidth]{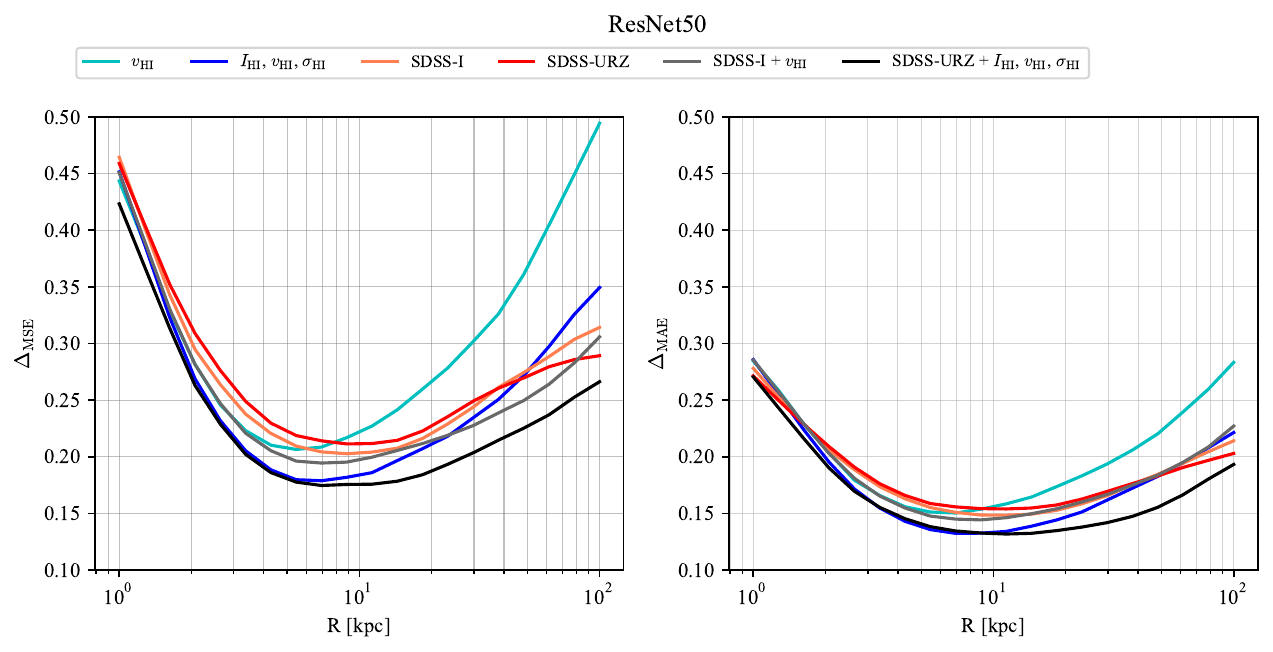}
    \caption{Root mean squared error (left) and mean absolute error (right) obtained for different sets of observations in the case of ResNet50}
    \label{fig:results_ResNet50}
\end{figure*}

We provide in figures \ref{fig:results_modelA}, \ref{fig:results_modelC} and \ref{fig:results_ResNet50} the comparisons of performance obtained with different combinations of observational inputs for the studied network architectures.

\newpage
\section{Network comparison (for a fixed dataset)}
\label{app:architecture_comparison}

\begin{figure*}
    \centering
    \includegraphics[width=\linewidth]{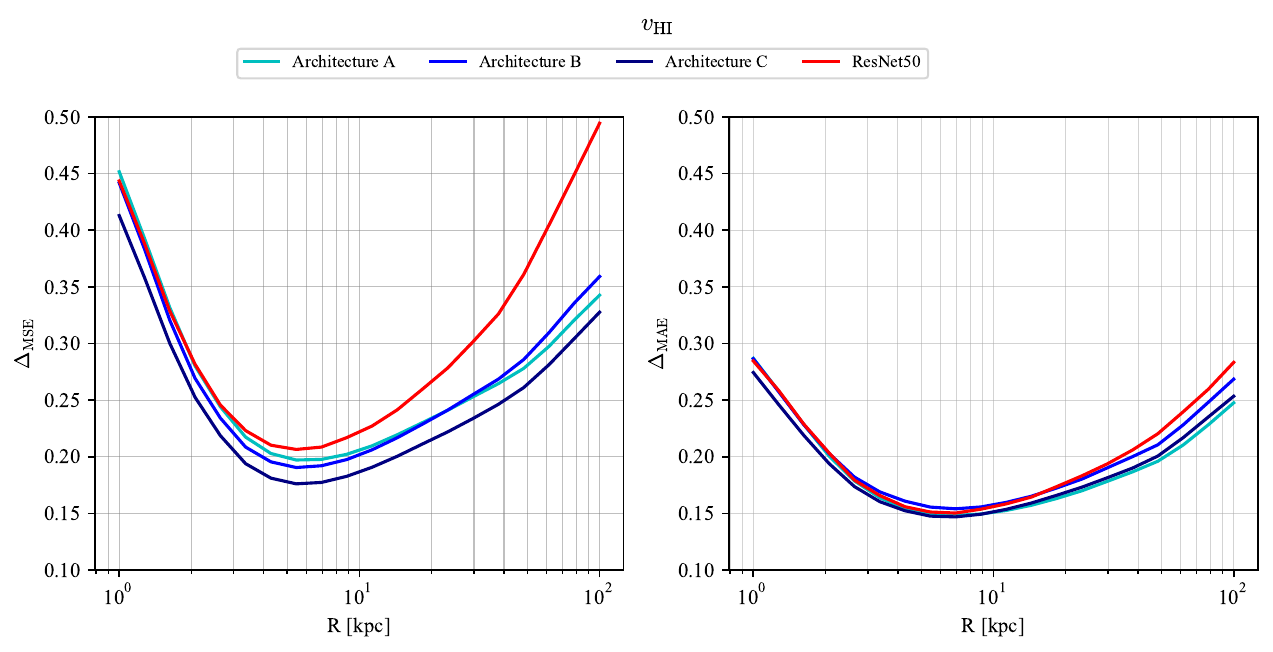}
    \caption{Root mean squared error (left) and mean absolute error (right) obtained for different neural network architectures when trained on HI line-of-sight velocity maps}
    \label{fig:results_HI1}
\end{figure*}

\begin{figure*}
    \centering
    \includegraphics[width=\linewidth]{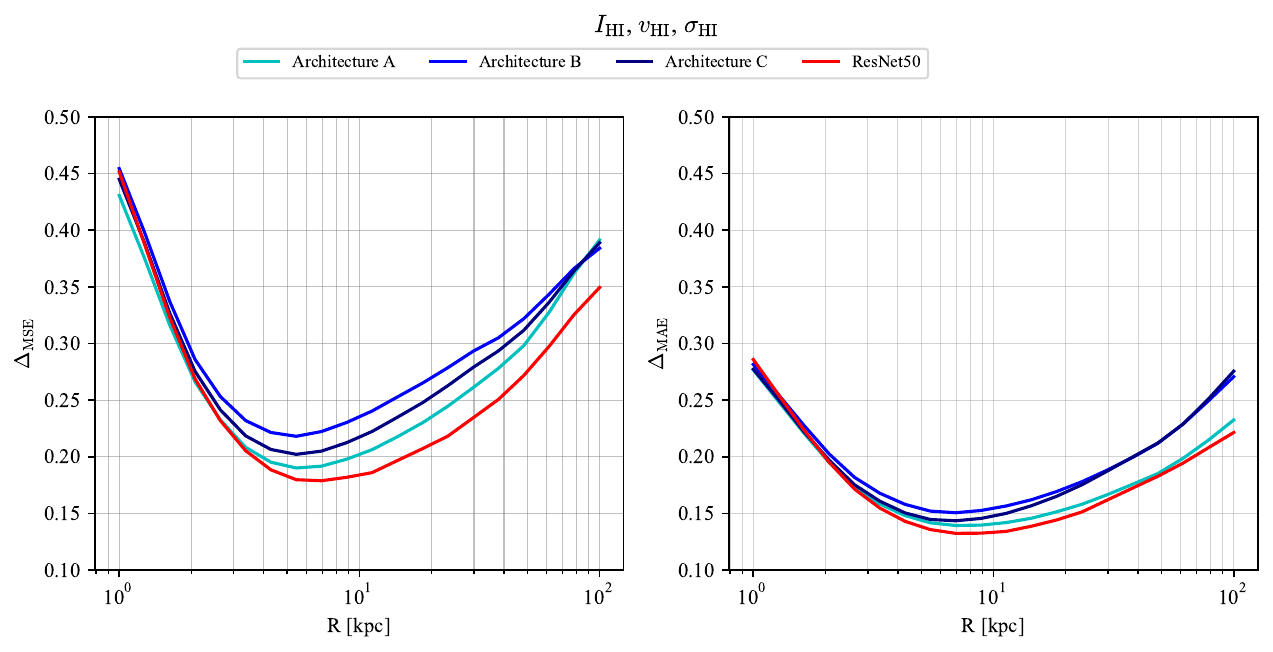}
    \caption{Root mean squared error (left) and mean absolute error (right) obtained for different neural network architectures when trained on HI intensity, line-of-sight velocity and line-of-sight velocity dispersion maps}
    \label{fig:results_HI012}
\end{figure*}

\begin{figure*}
    \centering
    \includegraphics[width=\linewidth]{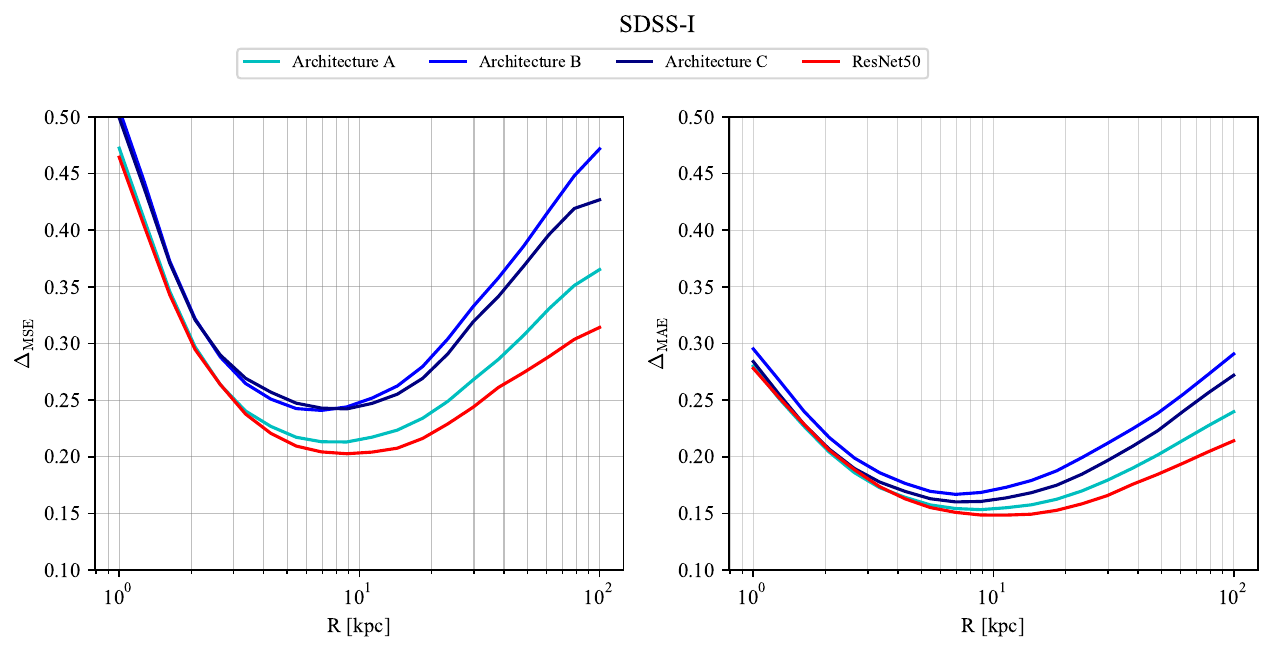}
    \caption{Root mean squared error (left) and mean absolute error (right) obtained for different neural network architectures when trained on SDSS I-band images}
    \label{fig:results_SDSS-I}
\end{figure*}

\begin{figure*}
    \centering
    \includegraphics[width=\linewidth]{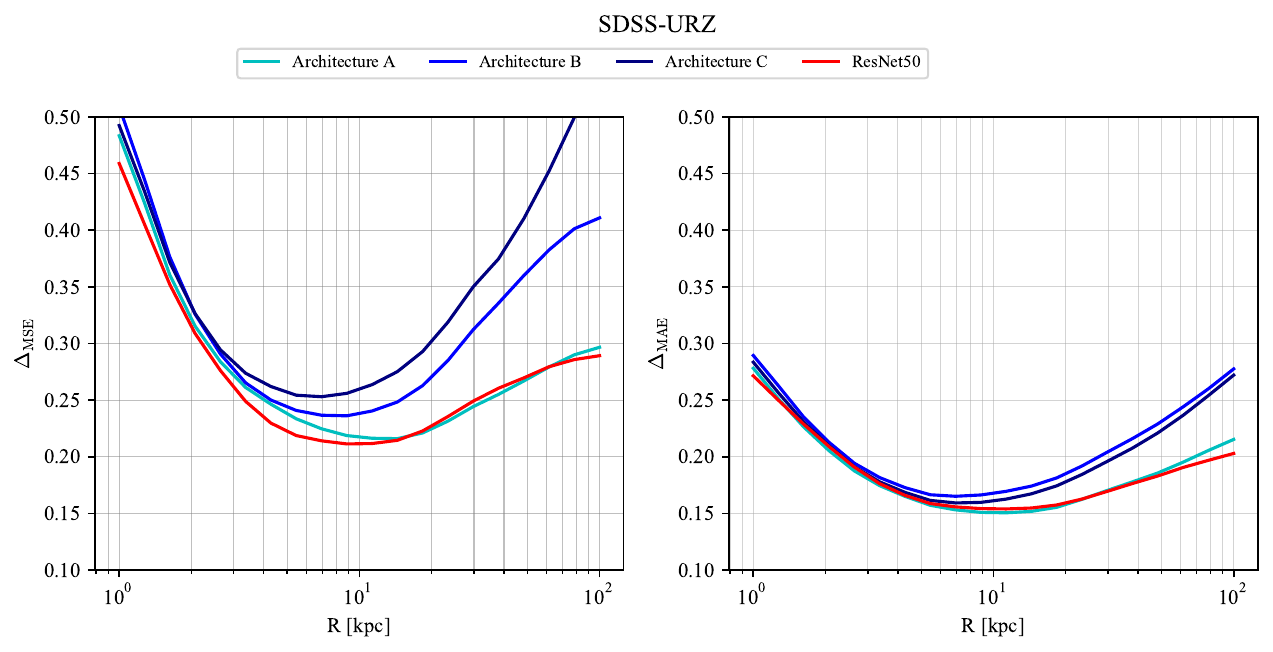}
    \caption{Root mean squared error (left) and mean absolute error (right) obtained for different neural network architectures when trained on SDSS U, R and Z-band images}
    \label{fig:results_SDSS-URZ}
\end{figure*}

\begin{figure*}
    \centering
    \includegraphics[width=\linewidth]{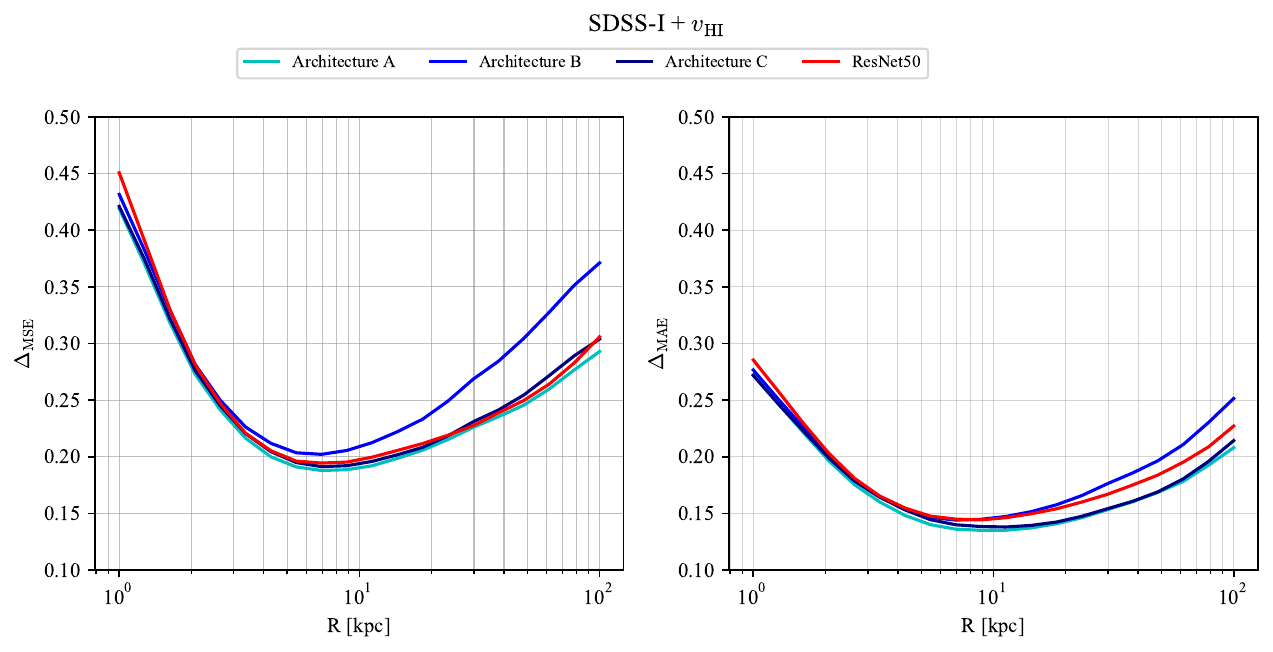}
    \caption{Root mean squared error (left) and mean absolute error (right) obtained for different neural network architectures when trained on HI line-of-sight velocity maps and SDSS I-band images}
    \label{fig:results_HI1+SDSS-I}
\end{figure*}

\begin{figure*}
    \centering
    \includegraphics[width=\linewidth]{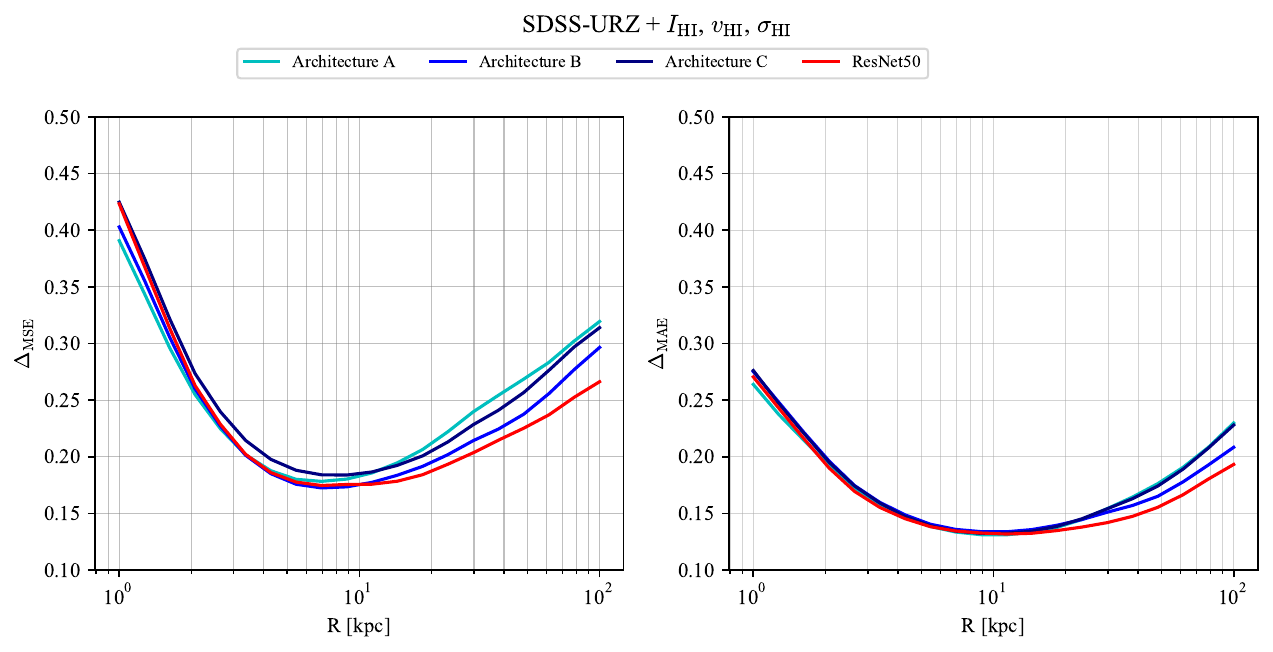}
    \caption{Root mean squared error (left) and mean absolute error (right) obtained for different neural network architectures when trained on HI intensity, line-of-sight velocity and line-of-sight velocity dispersion maps as well as SDSS U, R and Z-band images}
    \label{fig:results_HI012+SDSS-URZ}
\end{figure*}

We provide in  \cref{fig:results_HI1,fig:results_HI012,fig:results_SDSS-I,fig:results_SDSS-URZ,fig:results_HI1+SDSS-I,fig:results_HI012+SDSS-URZ} comparisons of performance obtained with different network architectures for a given combination of observational inputs.

\newpage
\section{Network description}
\label{app:network_description}

We provide in tables \ref{tab:modelA}, \ref{tab:modelB} and \ref{tab:modelC} the summary of the architectures of custom networks used in our work. For the structure of ResNet50 see~\cite{resnet, chollet2015keras}.

\begin{table*}
	\centering
	\begin{tabular}{c|c}
		Layer & Details \\ \hline \hline
		2D convolution & 64 kernels, $5 \times 5$ px kernel size, 2 px stride, ReLU activation \\
		2D max pooling & $2 \time 2$ px pooling \\
		Dropout & 25\% dropout fraction \\
		Batch normalization & \\ \hline
		2D convolution & 128 kernels, $5 \times 5$ px kernel size, 2 px stride, ReLU activation \\
		2D max pooling & $2 \time 2$ px pooling \\
		Dropout & 25\% dropout fraction \\
		Batch normalization & \\ \hline
		2D convolution & 128 kernels, $5 \times 5$ px kernel size, 2 px stride, ReLU activation \\
		2D max pooling & $2 \time 2$ px pooling \\
		Dropout & 25\% dropout fraction \\
		Batch normalization & \\ \hline
		Dense & 256 units, ReLU activation \\
		Dropout & 25\% dropout fraction \\
		Batch normalization & \\ \hline
		Dense & 128 units, ReLU activation \\ 
		Dropout & 25\% dropout fraction \\
		Batch normalization & \\ \hline
		Dense & 64 units, ReLU activation \\
		Dropout & 25\% dropout fraction \\
		Batch normalization & \\ \hline
		Dense (output) & 20 units, linear activation
	\end{tabular}
	\caption{The structure of network architecture A. The network has in total $\sim 7 \cdot 10^5$ trainable weights (exact number depends on the number of input channels).}
	\label{tab:modelA}
\end{table*}

\begin{table*}
	\centering
	\begin{tabular}{c|c}
		Layer & Details \\ \hline \hline
		2D max pooling & $2 \time 2$ px pooling \\
		Dropout & 50\% dropout fraction \\
		Batch normalization & \\ \hline
		2D convolution & 128 kernels, $5 \times 5$ px kernel size, 2 px stride, ReLU activation \\
		2D max pooling & $2 \time 2$ px pooling \\
		Dropout & 50\% dropout fraction \\
		Batch normalization & \\ \hline
		2D convolution & 256 kernels, $5 \times 5$ px kernel size, 2 px stride, ReLU activation \\
		Batch normalization & \\ \hline
		Dense & 256 units, ReLU activation \\
		Dropout & 50\% dropout fraction \\
		Batch normalization & \\ \hline
		Dense & 128 units, ReLU activation \\ 
		Dropout & 50\% dropout fraction \\
		Batch normalization & \\ \hline
		Dense & 64 units, ReLU activation \\
		Dropout & 50\% dropout fraction \\
		Batch normalization & \\ \hline
		Dense (output) & 20 units, linear activation
	\end{tabular}
	\caption{The structure of network architecture B. The network has in total $\sim 1.3 \cdot 10^6$ trainable weights (exact number depends on the number of input channels).}
	\label{tab:modelB}
\end{table*}

\begin{table*}
	\centering
	\begin{tabular}{c|c}
		Layer & Details \\ \hline \hline
		2D convolution & 64 kernels, $5 \times 5$ px kernel size, 2 px stride, ReLU activation \\
		2D max pooling & $2 \time 2$ px pooling \\
		Dropout & 50\% dropout fraction \\
		Batch normalization & \\ \hline
		2D convolution & 128 kernels, $5 \times 5$ px kernel size, 2 px stride, ReLU activation \\
		Dropout & 50\% dropout fraction \\
		Batch normalization & \\ \hline
		2D convolution & 256 kernels, $5 \times 5$ px kernel size, 2 px stride, ReLU activation \\
		Dropout & 50\% dropout fraction \\
		Batch normalization & \\ \hline
		2D convolution & 256 kernels, $3 \times 3$ px kernel size, 1 px stride, ReLU activation \\
		Dropout & 50\% dropout fraction \\
		Batch normalization & \\ \hline
		Dense & 256 units, ReLU activation \\
		Dropout & 50\% dropout fraction \\
		Batch normalization & \\ \hline
		Dense & 64 units, ReLU activation \\
		Dropout & 50\% dropout fraction \\
		Batch normalization & \\ \hline
		Dense (output) & 20 units, linear activation
	\end{tabular}
	\caption{The structure of network architecture C. The network has in total $\sim 2.2 \cdot 10^6$ trainable weights (exact number depends on the number of input channels).}
	\label{tab:modelC}
\end{table*}


\bsp	
\label{lastpage}
\end{document}